\documentclass[a4paper,11pt]{article}
\usepackage{jheppub}
\usepackage[T1]{fontenc}
\usepackage{amsmath,amssymb}
\pdfoutput=1
\usepackage[utf8x]{inputenc}
\usepackage{graphicx,caption}
\usepackage{epstopdf}
\usepackage{subfigure}
\usepackage{footnote}
\usepackage[export]{adjustbox}
\usepackage{cancel}
\DeclareGraphicsExtensions{.png,.eps,.pdf}
\renewcommand{\baselinestretch}{1.5}

\def\beq{\begin{equation}}
\def\eeq{\end{equation}}
\def\barr{\begin{array}}
\def\earr{\end{array}}
\def\dis{\displaystyle}
\def\tev{\, {\rm TeV}}
\def\gev{\, {\rm GeV}}
\def\fb{\, {\rm fb}}

\usepackage[normalem]{ulem}

\definecolor{darkgreen}{cmyk}{1,0,1,0.4}

\def\lapp{\mathrel{\rlap{\raise.5ex\hbox{$<$}}
                    {\lower.5ex\hbox{$\sim$}}}}
\def\gapp{\mathrel{\rlap{\raise.5ex\hbox{$>$}}
                    {\lower.5ex\hbox{$\sim$}}}}
\def\intey{\int_{0}^{\pi} dx_4}
\def\intez{\int_{0}^{\pi} dx_5}
\def\be{\begin{eqnarray*}}
\def\ee{\end{eqnarray*}}
\newcommand\summ{\sum_{n > 0}}

\title{Living Orthogonally: Quasi-universal Extra Dimensions}

\author[a]{Mathew Thomas Arun}
\affiliation[a]{Center for High Energy Physics, Indian Institute of Science, Bangalore-560 012, India}
\author[b]{Debajyoti Choudhury}
\author[b]{Divya Sachdeva}
 \emailAdd{divyasachdeva951@gmail.com}
\affiliation[b]{Department of Physics and Astrophysics, University of Delhi, Delhi 110 007, India}

\abstract{ The minimal Universal Extra Dimension scenario is highly
  constrained owing to opposing constraints from the observed relic
  density on the one hand, and the non-observation of new states at
  the LHC on the other. Simple extensions in five-dimensions can only
  postpone the inevitable.  Here, we propose a six-dimensional
  alternative with the key feature being that the SM quarks and
  leptons are localized on orthogonal directions whereas gauge bosons
  traverse the entire bulk. Several different realizations of electroweak
  symmetry breaking are possible, while maintaining agreement with low 
  energy observables. This model is not only consistent with all the
  current constraints opposing the minimal Universal Extra Dimension
  scenario but also allows for a multi-TeV dark matter particle
  without the need for any fine-tuning.  In addition, it promises a
  plethora of new signatures at the LHC and other future experiments.}

\begin{document}
\maketitle
\section{Introduction}
    \label{sec:intro}

Although the spectacular discovery of the long-sought for Higgs
boson~\cite{Aad:2012tfa,Aad:2012an,Chatrchyan:2012xdj} is cited as the
completion and the latest vindication of the Standard Model (SM),
certain questions remain unanswered. These pertain to the existence of
Dark Matter (DM), the origin of the baryon asymmetry in the Universe,
the existence of multiple generations of fermions, the hierarchy in
fermion masses and mixing, and, last but not the least, the stability
of the Higgs sector under quantum corrections. The pursuit of answers
to such questions has led to two different paradigms for the
exploration of physics beyond the SM. The top to bottom approach
posits a UV complete model, usually motivated to solve one or more of
the outstanding problems (including the hierarchy), and delves into its
``low-energy'' consequences.  Unfortunately, the most straightforward
of them, whether it be supersymmetry or warped extra dimensions are
faced with stringent constraints from observations. Hence, one is
forced to consider non-minimal versions, an exercise that could,
potentially, be non-intuitive on account of the lack of clear
principles. The second, or bottom to top approach envisages simplified
models, which while not addressing the UV scale physics, can explain
certain anomalies in particle physics experiments (including but not
limited to those at colliders) and possibly also cosmological
observations.

Over the years, several attempts have been made to address the aforementioned questions, albeit only with partial
success. One such stream of thought envisages a world in more than
three space dimensions as a possible panacea to some of the ills of
the SM, and, in this paper, we concentrate on this possibility. A
particularly simplistic version was given the name Minimal Universal
Extra Dimensions (mUED), wherein the SM is extended to propagate in a
5 dimensional space-time orbifolded to $M^4 \otimes S^1/Z_2$, with
$M^4$ being the four-dimensional space-time with Lorentz symmetry. If
the radius of the fifth dimension be small enough, this leaves us with
an effective 4-dimensional theory with every SM particle having a
corresponding infinite tower of Kaluza Klein (KK) modes separated at
equal energy gaps related to the inverse of the compactification
scale.

Being an non-renormalizable model, mUED is considered as an effective
theory valid up to a cut-off $\Lambda$, typically assumed to be 5--40
times the inverse radius. One of the main attractions of MUED was the
fact that, while the conservation of the KK-number (or, in other
words, the momentum in the compactified direction) is broken by
quantum corrections, a $Z_2$ symmetry, called KK-parity, is
maintained nevertheless.  This has the consequence that the lightest
of the level-1 KK partners (upon accounting for the radiative effects
\cite{Cheng:2002iz}, normally the cousin of the hypercharge gauge
boson) is absolutely stable, thereby being a natural candidate for the
Dark Matter
particle~\cite{Servant:2002aq,Kong:2005hn,Burnell:2005hm,Kakizaki:2006dz,Belanger:2010yx}. This
model has been studied in great detail, since it bore resemblance with
the LHC signatures of the minimal supersymmetric standard model, except for the spin of the individual
particles.

While the electroweak~\cite{Baak:2011ze} and
flavour~\cite{Haisch:2007vb,Dey:2016cve} observables impose only
relatively weak bounds on the compactification scale, namely $R^{-1}
  \gtrsim 750 \gev$ and $R^{-1} \gtrsim 600 \gev$
respectively, the
current LHC bounds are much stronger, and emanate from the study of a
variety of final
states~\cite{Choudhury:2016tff,Deutschmann:2017bth,Beuria:2017jez},
such as multijets, or dileptons with jets, each accompanied by missing
transverse energy ($\cancel E_T$), owing to the presence of the lightest
KK-particle. With $\Lambda R$ ($R$ being the compactification scale)
determining the mass splittings, the detection efficiency (for a given
$R^{-1}$) and, hence, the experimental reach is also determined by this. The
excellent agreement of the ATLAS data (at 13 TeV) on
dileptons~\cite{ATLAS:2016xcm} and multijets with a single-lepton
\cite{ATLAS:2016lsr} with the SM expectations, serves to exclude
$R^{-1} \lesssim 1400 \gev$ (for $\Lambda R \gtrsim 10$) or $R^{-1}
\lesssim 1500 \gev$ (for $\Lambda R \lesssim 10$) at 95\% C.L.

On the other hand, the agreement of the consequent relic density with
the WMAP\cite{Komatsu:2010fb} or the Planck~\cite{Ade:2015xua} data
demands that $1250 \gev \lesssim R^{-1} \lesssim 1500 \gev$
\cite{Belanger:2010yx}. These two sets of results are
in serious conflict with
each other, and continuing validity of this model would require
substantial alterations.  The simplest solution, of course, would be
to break KK-parity and give up on the DM-candidate. A more attractive
proposition would be to somehow alter the spectrum so as to either
suppress the relic abundance or relax the LHC constraints by raising
the masses of the strongly interacting particles. A partial solution
can be achieved by invoking perfectly-tuned brane localized and/or
higher dimensional terms in the Lagrangian.  While the existence of
such terms is anticipated as the UED is only an effective field theory
with a cutoff $\Lambda$, the symmetric character of such terms
(necessary for stability of the DM) is to be assured by imposing a
KK-parity. Attractive as this proposition might be, it only offers a
partial solution as the twin constraints (LHC and relic density) would
require rather large sizes for these terms, that, ostensibly, arise
from quantum corrections.

While for reasons of simplicity (and on
  account of the constraints from electroweak observables being
  relatively mild in this case), most models have concentrated on a
five-dimensional world, this, though, need not be the case. Indeed,
the extension to six dimensions~\cite{Burdman:2005sr,Dobrescu:2004zi,
  Maru:2009wu, Cacciapaglia:2009pa, Dohi:2010vc} brings forth its own
advantages, {\em e.g.,} in the context of explaining the existence of
three fermion generations, or the reconciliation with the
non-observance of proton decay. On the other hand, the simplest
generalization, typically, results in even stronger constraints from
relic density~\cite{Dobrescu:2007ec}. Thus, smarter generalizations
are called for. Embedding the UED in a 6-dimensional warped space
\cite{Arun:2016csq,Arun:2017zap}, for example, makes it
possible to evade the relic density bounds by exploiting the
$s$-channel annihilation channel with graviton mediated by a
KK-graviton exchange.  In this paper we offer a different solution,
one that leads to a very rich collider phenomenology.

The rest of this article is constructed as follows. We begin by
setting up the formalism, followed by details of the model including
the breaking of the electroweak symmetry and culminating with the
Feynman rules.  This is followed (in Sec. \ref{sec:spectrum}) by an
examination of the mass-splittings wrought by quantum corrections.  A
detailed study of the relic abundance and the consequent constraints
imposed on the parameter space of the theory is presented in Section
\ref{sec:dark_constraints}. Also discussed are the prospects for
direct and indirect detection experiments.  An independent set of
constraints arise from new particle searches at the LHC and these are
discussed in Section \ref{sec:collider}.  Finally, we conclude in
Section \ref{sec:concl} and list some open questions.

\section{The Model}
    \label{sec:model}
Consider a 6-dimensional flat space-time orbifolded on $M^4\otimes
(S^1/Z_2) \otimes (S^1/Z_2)$ parametrized\footnote{Note that
    this can also be thought of as $M^4 \otimes T^2 / (Z_2 \otimes
    Z_2)$, with the two radii of the $T^2$ being distinct.} by the
coordinates ($x_{\mu},x_4,x_5$), where $M^4$ is the 4-dimensional
space ($x_{\mu}$) that obeys Lorentz symmetry, and $x_{4,5}$ are
compact and dimensionless. The line element for this space-time is
given as,
\[
ds^2 = \eta_{\mu \nu}dx^{\mu}dx^{\nu} + R_q^2 dx_4^2 + r_\ell^2 dx_5^2 \ ,
\]
where $R_q$ and $r_\ell$ are the compactification radii in the $x_4$
and $x_5$ directions respectively. The latter are orbifolded by
individual $Z_2$'s, and periodic boundary conditions ($x_4 \rightarrow
x_4+ \pi $ and $x_5 \rightarrow x_5+ \pi$) are assumed. This
particular orbifolding demands the sides of the six-dimensional space
to be protected by 4-branes (or 4+1 dimensional hyper surfaces). This
feature is distinctively different from the toroidal orbifolding
($T^2/Z_2$) where the branes are present only at the corners and are
co-dimension 2. It should be noted that it is 
the mutual independence of the two orbifoldings that allows for $R_l\neq r_q$,
rendering the six-dimensional bulk spacetime to be 
different (similar spacetime has also been 
considered in Refs.\cite{Li:2001qs,Li:2002xw})
from the chiral square~\cite{Dobrescu:2004zi}. 

The key feature is that the SM quarks and leptons are extended not to
the entire bulk, but only to orthogonal directions. Apart from other
consequences (to be elaborated on later) this immediately decouples
the inter-level mass splittings for quarks and leptons, thereby
raising the possibility that the LHC bounds could be evaded while
satisfying the constraints from relic abundance.  To be specific,
we consider the case where the lepton fields exist
  only on the 4-brane at $x_4=0$ while quarks exist, symmetrically, on
  a different set of 4-branes, at $x_5 = 0$ and $x_5 = \pi$, that are
  orthogonal to the leptonic brane. This difference
  in the assignments of quarks and leptons has very important
  ramifications. As we shall see later, the symmetric assignation of
  the quark field will allow us to retain the $Z_2$ symmetry
  associated with the ``leptonic direction'', viz.  $x_5 \to \pi -
  x_5$ , whereas the corresponding symmetry in the $x_4$ direction is
  manifestly lost.  To remind us of this aspect (one which will have
immense consequences), we will denote the remaining discrete symmetry of the
model by $Z_2^{\ell} \otimes \cancel{Z}_2^q$ (where the slash denotes
the explicit breaking). The free Lagrangian\footnote{The
    promotion to a gauge-invariant version is straightforward and is
    described later.} for the fermions is, then, given by
\beq
\barr{rcl}
S_{\rm fermion} &=& \dis \int \int_{0}^{\pi} \int_{0}^{\pi}
dx^{\mu} dx_4 dx_5  \,
\left[\sqrt{g_1} {\cal L}_q \, \frac{\left\{\delta(x_5) + \delta(x_5-\pi )\right\}}{2}
  +  \sqrt{g_2}{\cal L}_\ell \, \delta(x_4)
     \right]
\\[2ex]
{\cal L}_\ell & \equiv & \dis
\bar{L}(-i \gamma^\mu \partial_\mu -i \gamma_5 r_\ell^{-1} \partial_5) L + \bar{E}(-i\gamma^\mu \partial_\mu -i \gamma_5 r_\ell^{-1}\partial_5) E 
\\[2ex]
  {\cal L}_q& \equiv& \dis \bar{Q}(-i\gamma^\mu \partial_\mu -i \gamma_4 R_q^{-1}\,
  \partial_4) Q
         + \bar{U}(-i\gamma^\mu \partial_\mu -i \gamma_4 R_q^{-1}\,\partial_4) U 
 \\
 &+&\dis
 \bar{D}(-i\gamma^\mu \partial_\mu -i \gamma_4 R_q^{-1}\,\partial_4) D \
\earr
\eeq
where $L/Q$ are the lepton/quark doublets and $E/U/D$ are the
lepton/quark singlets and $\sqrt{g_1} = R_q$, $\sqrt{g_2} = r_\ell$.
The fermions, being 5-dimensional\footnote{Consequently, both
    $\gamma_{4,5}$ can be thought of as being identical to the usual
    $\gamma_5$ in a four-dimensional theory.}, are vector like. The
unwanted zero-mode chiral states can be projected out by orbifolding
appropriately, viz., for $x_5 \to - x_5$,
\[
L(x_\mu,x_5) = \gamma^5  L(x_\mu,-x_5) \ ,
              \hspace{0.5em} E(x_\mu,x_5) = - \gamma^5  E(x_\mu,-x_5) 
\]
while for $x_4\rightarrow-x_4$, 
\[
 Q(x_\mu,x_4) = \gamma^5  Q(x_\mu,-x_4) \ ,
     \hspace{0.5em} U(x_\mu,x_5) = - \gamma^5  U(x_\mu,-x_5) \ ,
     \hspace{0.5em} D(x_\mu,x_5) = - \gamma^5  D(x_\mu,-x_5) \ .
\]
The corresponding Fourier decompositions of the fermion fields are
given by
\beq
\barr{rcl}
\sqrt{\pi r_\ell} \,L(x_{\mu},x_5) & = & \dis L_l^{(0,0)}(x_{\mu}) 
+ \sqrt{2} \; \sum_{p > 0} 
   \Big(L_l^{(0,p)}(x_{\mu}) \cos\left(p x_5\right) 
        + L_r^{(0,p)}(x_{\mu}) \sin\left(p x_5\right)\Big) 
\\[1.5ex]
\sqrt{\pi r_\ell} \, E(x_{\mu},x_5) & = & \dis  E_r^{(0,0)}(x_{\mu}) 
+ \sqrt{2} \; \sum_{p > 0} 
   \Big(E_l^{(0,p)}(x_{\mu}) \cos\left(p x_5\right) 
        + E_r^{(0,p)}(x_{\mu}) \sin\left(p x_5\right)\Big) 
\\[1.5ex]
\sqrt{\pi R_q} \, Q (x_{\mu},x_4) & = & \dis  Q_l^{(0,0)}(x_{\mu}) 
+ \sqrt{2} \; \sum_{n > 0} 
      \Big(Q_l^{(n,0)}(x_{\mu}) \cos\left(n x_4\right) 
          + Q_r^{(n,0)}(x_{\mu}) \sin\left(n x_4\right)\Big) 
\\[1.5ex]
\sqrt{\pi R_q} \, U(x_{\mu},x_4) & = & \dis  U_r^{(0,0)}(x_{\mu}) 
+ \sqrt{2}\; \sum_{n > 0} 
      \Big(U_l^{(n,0)}(x_{\mu}) \cos\left(n x_4\right) 
          + U_r^{(n,0)}(x_{\mu}) \sin\left(n x_4\right)\Big) 
\\
\sqrt{\pi R_q} \, D(x_{\mu},x_4) & = & \dis  D_r^{(0,0)}(x_{\mu}) 
+ \sqrt{2} \; \sum_{n > 0} 
      \Big(D_l^{(n,0)}(x_{\mu}) \cos\left(n x_4\right) 
          + D_r^{(n,0)}(x_{\mu}) \sin\left(n x_4\right)\Big) \ .
\earr
   \label{fermion_mode_expn}
\eeq
As expected, both the leptons and quarks have a single tower each. The
former has only the $(0,p)$ modes with masses being given by
$m_\ell^{(0,p)} = p \, r_\ell^{-1}$ where we have neglected the SM
mass for the zero mode. The quarks, on the other hand, possess only
the $(n,0)$ modes with masses $m_q^{(n,0)} = n \, R_q^{-1}$. Naively,
one would expect that we would need $R_q^{-1}$ to be sufficiently
larger than $r_\ell^{-1}$. However, as we shall see later, this
requirement is not strictly true.

With quarks and leptons being extended in different directions, it is
obvious that at least the electroweak gauge bosons must traverse the
entire six-dimensional bulk\footnote{An alternative would be to
  consider separate electroweak gauge groups for the two sectors,
  confining each to the corresponding branes. This could be broken
  down to the diagonal subgroup (to be identified with the SM
  symmetry) through appropriate
  Higgses~\cite{Georgi:1989ic,Georgi:1989xz,Choudhury:1991if}. We
  eschew this path.}.  Before we delve into this, we offer a quick
recount of the gauge sector in a generic higher-dimensional model.

\subsection{Gauge Bosons: A lightning review}

Naively, a five-dimensional gauge field, on compactification, should
decompose into a four-dimensional vector field and an adjoint scalar,
whereas for a six-dimensional field, there should be two such scalars
instead. However, a simple counting of the degrees of freedom
(especially for the KK-levels), shows immediately that, in each case,
one of the scalar modes must vanish identically.

\begin{table}[!ht]
\begin{center}
\begin{tabular}{|c|c|}
\hline
$x_4 \rightarrow - x_4, \quad 
x_5 \rightarrow x_5$ & $x_4 \rightarrow x_4, \quad x_5 \rightarrow - x_5$ \\
\hline
$A_{\mu}(x^{\mu},x_{4},x_{5}) = A_{\mu}(x^{\mu},-x_{4},x_{5})$ & $A_{\mu}(x^{\mu},x_{4},x_{5}) =  \dis A_{\mu}(x^{\mu},x_{4},-x_{5})$\\
$A_{4}(x^{\mu},x_{4},x_{5})  = -A_{4}(x^{\mu},-x_{4},x_{5})$ & $A_{4}(x^{\mu},x_{4},x_{5}) = A_{4}(x^{\mu},x_{4},-x_{5})$\\
$A_{5}(x^{\mu},x_{4},x_{5})  =  A_{5}(x^{\mu},-x_{4},x_{5})$ & $A_{5}(x^{\mu},x_{4},x_{5}) =  -A_{5}(x^{\mu},x_{4},-x_{5})$\\
$\Theta(x^{\mu},x_{4},x_{5})  = \Theta(x^{\mu},-x_{4},x_{5})$ & $\Theta(x^{\mu},x_{4},x_{5}) = \Theta(x^{\mu},x_{4},-x_{5})$ \\
\hline
\end{tabular}
\vskip 2ex
\caption{\em Boundary conditions for the gauge fields.}
\label{tab:gaugebc}
\end{center}

\end{table}

To begin with, we look at the simpler abelian case and, then, graduate
to the Standard Model gauge content.  The action is given by
\beq
 S=\int d^{4}x\intey\intez \, \sqrt{g} \left(-\frac{1}{4}F_{MN}F^{MN} 
    + {\cal L}_{gf} \right) \ ,
\eeq
where $\sqrt{g} = R_q r_\ell$ and $M, N = 0,\dots,5$.  To eliminate the spurious degrees of
freedom corresponding to the gauge freedom, viz,
\[
A_{M}(x^{N})\rightarrow A'_{M}(x^{N})=A_{M}(x^{N})+\partial_{M}\Theta(x^{N}) \ ,
\]
we choose to work with the generalized $ R_\xi $ gauge and
\beq
{\cal L}_{gf}=-\frac{1}{2\xi}\left[\partial_{\mu}A^{\mu}
     -\xi \, (R_q^{-1}\partial_{4}A_{4}+r_{\ell}^{-1}\partial_{5}A_{5}) \right]^{2} \ .
\eeq
This has the further advantage of eliminating terms connecting $A_\mu$
to $A_{4,5}$. To compactify on the orbifold, we need to impose
boundary conditions which are summarized in Table~\ref{tab:gaugebc}.
We may, now, effect mode decomposition, viz.
\beq
\barr{rcl}
\pi \, \sqrt{R_q \, r_\ell}\, A_{\mu}(x^{\mu},x_{4},x_{5})& = & \dis  A^{(0,0)}_{\mu}(x^{\mu}) 
   + \sqrt{2} \, \sum_{j>0} A^{(j,0)}_{\mu}(x^{\mu})\;
          \cos\left(j x_{4} \right)
\\
         & + & \dis \sqrt{2} \sum_{k> 0} A^{(0,k)}_{\mu}(x^{\mu})\;
          \cos\left(k x_{5} \right) 
   + 2 \, \sum_{j,k > 0} A^{(j,k)}_{\mu}(x^{\mu}) \;
          \cos\left(j x_{4}\right) \cos\left(k x_{5}\right) 
\\[1.0ex]
\pi \, \sqrt{R_q \, r_\ell} \, A_{4}(x^{\mu},x_{4},x_{5})& = & \dis  
\sqrt{2} \, \sum_{j> 0}A^{(j,0)}_{4}(x^{\mu})\;
          \sin\left(jx_{4}\right)
+ 2\, \sum_{j,k > 0} A^{(j,k)}_{4}(x^{\mu}) \, 
          \sin\left(jx_{4}\right)\cos\left(kx_{5}\right) 
\\[1.0ex]
\pi \, \sqrt{R_q \, r_\ell} \, A_{5}(x^{\mu},x_{4},x_{5})& = & \dis 
\sqrt{2} \, \sum_{k> 0} A^{(0,k)}_{5}(x^{\mu})\;
          \sin\left(k x_{5} \right)
+ 2 \, \sum_{j,k > 0} A^{(j,k)}_{5}(x^{\mu}) \, 
          \cos\left(j x_{4}\right)\sin\left(kx_{5} \right)\ .
\earr
\label{gaugekkdecomposition}
\eeq
The factor $\pi \, \sqrt{R_q r_\ell}$ serves to maintain canonical
commutation relations for the KK components on compactification down
to four dimensions. Clearly, $A_{4,5}^{(j,k)}$ transform as
four-dimensional Lorentz scalars. Although, for $j,k \neq 0$, the
kinetic terms mix the fields, these can be diagonalized provided we
redefine them as
\beq
\barr{rcl}
V^{(j,k)}_{1} & = & \dis \frac{1}{M_{j,k}} \,
     \left(\frac{k}{r_\ell}A^{(j,k)}_{4}-\frac{j}{R_q}A^{(j,k)}_{5}\right)\\
V^{(j,k)}_{2} & = & \dis \frac{1}{M_{j,k}}
    \left(\frac{j}{R_q}A^{(j,k)}_{4}+\frac{k}{r_\ell} A^{(j,k)}_{5}\right)
\earr
\label{adjointscalars}
\eeq where, as usual, $M_{j,k}^2 = j^2 / R_q^2 + k^2/r_\ell^2$.  Under
such a redefinition, after integrating out the extra dimensions, the
effective four-dimensional Lagrangian density is
\beq
\barr{rcl}
{\cal L} & = & \dis 
    \sum_{j,k} \left[\frac{-1}{4} \, F^{(j,k)}_{\mu\nu} \, F^{(j,k)\mu\nu}
                          +M^{2}_{j,k} \, A^{(j,k)}_{\mu} \, A^{(j,k)\mu} 
                         - \frac{1}{2\xi}(\partial_{\mu}A^{(j,k)\mu})^{2} \right]
\\
    & + & \dis \sum_{j,k} \left[(\partial_{\mu}V^{(j,k)}_{1})^{2}-M^{2}_{j,k}V^{(j,k) 2}_{1} + (\partial_{\mu}V^{(j,k)}_{2})^{2}-\xi M^{2}_{j,k}V^{(j,k) 2}_{2} \right]
\earr
\eeq
Using Eq~\ref{gaugekkdecomposition}, it is trivial to see that $V_1$
does not exist if $j=0$ or $k=0$. This reflects the orbifolding we
have in this geometry.  Thus, we have a double Kaluza-Klein tower for
a vector field along with a pair of charge-neutral scalars, with these
being degenerate at every level\footnote{The degeneracy would be
  lifted on the inclusion of the quantum corrections.}.  In the
unitary gauge ($\xi \to \infty$), the scalars $V_{2}^{(j,k)}$ become
non propagating and the only (real) scalar fields left are
$V_{1}^{(j,k)}$. This disappearance of one tower of the adjoint
scalars can be understood in terms of the appearance of the
longitudinal modes for the corresponding gauge boson levels.

\subsection{The complete field assignment}
The decomposition for non-abelian gauge boson is identical to that for
the abelian case discussed above, with the added complication of the
self interactions and the ghost fields needed to consistently define
the quantum theory. The gauge field is now expressible as $A_M = A_M^a
\, t^a$, and the field strength tensor as
\[
    F_{MN} = \partial_M \, A_N - \partial_N \, A_M + g \, [A_M, A_N]
\]
where $t^a$ are the generators and $g$ the six dimensional coupling
constant.  Once again, the gauge Lagrangian is written as
\[
   {\cal L}_{\rm kin} = \frac{-1}{4} \, tr(F_{MN} \, F^{MN}) \ ,
\]
apart from the gauge fixing and the ghost terms. For convenience, we
separate $F_{MN}$ into three sets $F_{\mu \nu}$, $F_{4 \mu}$, $F_{5
  \mu}$ and $F_{45}$ as in the abelian case, with the understanding
that each continues to be a function of all six dimensions. The
bilinear terms, on KK reduction give rise, in analogy with the abelian
case, to double towers of gauge bosons as well as of a pair of scalars
$V_{1,2}$ in the adjoint representation. Once again, in the unitary
gauge, $V_2$ decouples entirely. We desist from expanding the
Lagrangian further to include the ghosts etc, since that is not
germane to the issues under consideration.

Since we want to decouple the compactification scales for the
strongly-interacting particles from those having only electroweak
interactions, we consider the gluons to
  exist only on the very same branes where the
quarks are located.  Thus, the Lagrangian for the gauge sector is
given by
\beq
\barr{rcl}
{\cal L}_{Gauge}&=& \dis -\frac{1}{4} B_{MN} B^{MN} -\frac{1}{4} W^a_{MN} W^{aMN} 
- \, \frac{1}{4} G^\aleph_{\bar M \bar N} G^{\aleph \bar M \bar N} \frac{\left[ \delta(x_5) + \delta(x_5-\pi) \right]}{2} \\				
{\cal L}_{GaugeFix.} &=& \dis \frac{-1}{2\xi} \left ( \partial^\mu B_\mu - \xi (R_q^{-1}\partial_4 B_4 + r_{\ell}^{-1}\partial_5 B_5) \right )^2  \\  
                &- & \dis \frac{1}{2\xi} \left ( \partial^\mu W^a_\mu - \xi (R_q^{-1}\partial_4 W^a_4+r_{\ell}^{-1}\partial_5 W^a_5) \right )^2\\
 		&- & \dis \frac{1}{2\xi} \left\{ \partial^\mu G^\aleph_\mu - \xi (R_q^{-1}\partial_4 G^\aleph_4) \right\}^2 \frac{\left[\delta(x_5) + \delta(x_5-\pi)\right]}{2}   \ ,
\earr
\eeq
where $\bar M, \bar N = 0,\dots,4$, $a = 1,2,3$, $\aleph = 1 \dots 8$. While the mode decomposition for the electroweak gauge
bosons would be analogous to that presented in the preceding section
(we will discuss symmetry breaking shortly), that for the gluon is
simpler and is given by
\beq
\sqrt{\pi R_q} G_\mu^\aleph (x,x_4) = 
  G_\mu^{\aleph(0,0)}(x) + \sqrt{2} \summ G_\mu^{\aleph(n,0)}(x) \cos (n x_4) \ , 
\eeq
namely a single tower with masses given by $n / R_q$.  In the unitary
gauge, expectedly, no adjoint scalar remains.

Before delineating the Feynman rules for the gauge interactions,
  it is mandatory that we precisely define the covariant derivatives
  for the fermions. This follows in the usual manner, namely,
  \beq\partial_M \to D_M \equiv \partial_M - i \, g_1 \, B_M - i \, g_2
  \, W_M^a T^a - i g_3 \, G_M^\aleph T^\aleph,\eeq with the understanding
  that $G_5$ vanishes identically (since both quarks and gluons exist
  only on two particular branes). Further, it is convenient to define
the following symbols:
\beq
\barr{rclcl}
& & b_{\vec j}& \equiv & 2^{-(\delta_{j_{1},0}+\delta_{j_{2},0}+\delta_{j_{3},0}+1)/2}
\\
\delta_{l}(\vec j) & = &\delta_{j_{1},j_{2},j_{3}} & \equiv &  
\delta_{j_{1}+j_{3}+j_{2},0}+\delta_{j_{1}+j_{3}-j_{2},0}+\delta_{j_{1}-j_{2}-j_{3},0}
                 + \delta_{j_{1}+j_{2}-j_{3},0}
\\
\delta_{r}(\vec j) & = & \bar{\delta}_{j_{1},j_{2},j_{3}} & \equiv & 
           -\delta_{j_{1}+j_{3}+j_{2},0}-\delta_{j_{1}-j_{2}-j_{3},0}+\delta_{j_{1}-j_{2}+j_{3},0}+\delta_{j_{1}+j_{2}-j_{3},0}
\\          
& & b'_{\vec j, \vec k} & \equiv & 
2^{-2 - (\delta_{j_{1},0}+\delta_{j_{2},0}+\delta_{j_{3},0}+\delta_{j_{4},0}
         + \delta_{k_{1},0}+\delta_{k_{2},0}+\delta_{k_{3},0}+\delta_{k_{4},0})/2}
\\
& & \delta^{(1)}_{j_{1},j_{2},j_{3},j_{4}} & \equiv & 
\delta_{j_{1}+j_{2}+j_{3}+j_{4},0}+\delta_{j_{1}+j_{2}+j_{3}-j_{4},0}+\delta_{j_{1}+j_{2}-j_{3}+j_{4},0}+\delta_{j_{1}+j_{2}-j_{3}-j_{4},0}
\\[0ex]
& &  & + & \dis 
 \delta_{j_{1}-j_{2}+j_{3}+j_{4},0}+\delta_{j_{1}-j_{2}+j_{3}-j_{4},0}+\delta_{j_{1}-j_{2}-j_{3}+j_{4},0}+\delta_{j_{1}-j_{2}-j_{3}-j_{4},0}
\earr
\eeq
In terms of these, the gauge couplings can be expressed as in 
Tables~\ref{tab:bulk_interactions} \& \ref{tab:bulk_interactions2}.

\begin{table}
\begin{center}
\begin{tabular}{|c|c|}
\hline
$A_{\mu}^{(n_1,p_1)}Q^{(n_2,0)}_{l/r}Q^{(n_3,0)}_{l/r}$ & 
     $g \, b_{\vec n} \, \delta_{l/r}(\vec n) \, 
               2^{(1 - \delta_{p_1,0})/2}$, for even $p_1$ (0 otherwise)
\\
\hline
$A_{\mu}^{(n_1,p_1)}L^{(0,p_2)}_{l/r}L^{(0,p_3)}_{l/r}$ & 
$g \, 2^{(1 - \delta_{n_1,0})/2} \, b_{\vec p} \, \delta_{l/r}(\vec p)$ 
\\
\hline
$\partial_{\nu}A_{\mu}^{a(n_1,p_1)}A^{\nu b(n_2,p_2)}A^{\mu c(n_3,p_3)}$ & 
$ \dis \frac{g}{2} \, f^{abc} \, b_{\vec n} b_{\vec p} \, 
          \delta_{l}(\vec n) \, \delta_{l}(\vec p)$ 
\\
\hline
$A_{\mu}^{b(n_1,p_1)}A_{\nu}^{c(n_2,p_2)}A^{\mu d(n_3,p_3)}A^{\nu e(n_4,p_4)}$ & 
$\dis \frac{g^2}{4} \, f^{abc}f^{ade} \, b'_{\vec n, \vec p} \, 
          \delta^{(1)}_{n_{1},n_{2},n_{3},n_{4}} \, \delta^{(1)}_{p_{1},p_{2},p_{3},p_{4}}$
\\
\hline
$A_{\mu}^{(n_1,p_1)}A_{\mu}^{(n_2,p_2)} \ H \ H \ $ & $\frac{g^2}{2} \, [1+(-1)^{n_1+n_2}]\, 
                               [1+(-1)^{p_1+p_2}]$
\\
\hline
$\bar{Q}^{(n_1,0)} d_{r}^{(n_2,0)} \ H \ $ & $\lambda_d \frac{[1+(-1)^{n_1+n_2}]}{2}$ 
\\
\hline
$\bar{L}^{(0,p_1)} e_{r}^{(0,p_2)} \ H \ $ & $\lambda_e \frac{[1+(-1)^{p_1+p_2}]}{2}$
\\
\hline
\end{tabular}
\vskip 2ex
\caption{\em Interactions of bulk gauge boson fields, fermions and Higgs
  field. $f^{abc}$ are the structure constants. The Lorentz structures
  are as in the SM.  It should be noted that since the gluons are
  localized on the same brane as the quarks, they have only a
  single KK-tower and KK-number conserving interactions. Here we have considered Higgs
localized on the four corners.}
\label{tab:bulk_interactions}
\end{center}
\end{table}
\begin{table}
\begin{center}
\begin{tabular}{|c|c|}
\hline
$A_{\mu}^{(n_1,p_1)}A_{\mu}^{(n_2,p_2)} \ H^{(n_3,p_3)} \ H^{(n_4,p_4)} \ $ & $\frac{g^2}{4} \, b'_{\vec n, \vec p} \, 
          \delta^{(1)}_{n_{1},n_{2},n_{3},n_{4}} \, \delta^{(1)}_{p_{1},p_{2},p_{3},p_{4}}$
\\
\hline
$\bar{Q}^{(n_1,0)} d_r^{(n_2,0)} \ H^{(n_3,p_3)} \ $ & $\lambda_d \, b_{\vec n} \, \delta_{l}(\vec n) \, 2^{(1 - \delta_{p_1,0})/2}$, for even $p_1$ (0 otherwise)
\\
\hline
$\bar{L}^{(0,p_1)} e_r^{(0,p_2)} \ H^{(n_3,p_3)} \ $ & $\lambda_e \, 2^{(1 - \delta_{n_1,0})/2} \, b_{\vec p} \, \delta_{l}(\vec p)$
\\
\hline
\end{tabular}
\vskip 2ex
\caption{\em Interactions of bulk gauge boson and fermion with Higgs
  field in the bulk.}
\label{tab:bulk_interactions2}
\end{center}
\end{table}

\section{Electroweak Symmetry Breaking}

While the discussion so far has been rather straightforward, complications may 
arise when symmetry breaking is introduced. Several realizations of the Higgs
sector is possible, each with its own distinctive
consequences. We illustrate this now, beginning with what,
  naively, might seem the simplest choice, before graduating to one that
  not only is easier to work with, but also more viable
  phenomenologically.

\subsection{Higgs at the corners}
We begin with a particularly simple 
one, wherein the Higgs field is localized to 3-branes 
at the four corners of the rectangle that the compact space is.
The corresponding Lagrangian is, then, given by
\begin{equation}
{\cal L}_H = - \frac{1}{4}\left[\frac{1}{2}(D_\mu H)^{\dagger}(D^\mu H) 
  + V(H^\dagger H)\right]
     \,  \left[\delta(x_4)+\delta(x_4-\pi )\right] \, 
         \left[\delta(x_5)+\delta(x_5-\pi ) \right].
   \label{Higgs_lagr}
\end{equation}
A nontrivial vacuum expectation value for $H$ breaks the electroweak
symmetry down to $U(1)_{\rm em}$. The very presence of the $\delta$-functions 
in ${\cal L}_H$ has an
interesting consequence in that 
non-diagonal mass terms connecting the even and odd gauge boson modes, separately,
are engendered. 
As a result, the physical $W$-- and $Z$--bosons
will have an admixture of all the even KK-modes
\footnote{Analogously, the odd gauge boson modes too would encounter mass-mixings
amongst themselves. This sector, however, does not concern us.}. The mixings would be suppressed, though, by factors
of the order of $g^2 v^2/M_{KK}^2$
where $v$ is the electro-weak scale and $g$ the coupling constant.  
 
The Yukawa coupling is as usual, namely
\beq
{\cal L}_{\rm Yuk.} = \lambda_u \, \bar{Q} \, u_r \, (i\sigma^2 H^*) 
  + \lambda_d \, \bar{Q} \, d_r \, H  + \lambda_e \, \bar{L} \, e \, H \ .
\eeq
The particular localization of the Higgs (as in
eq.(\ref{Higgs_lagr})) implies that, of the various fermion
excitations, only the left-handed $SU(2)_L$-doublets and right-handed
singlets couple to the Higgs. The wavefunctions for the other (wrong)
chiralities vanish identically at the corners---see
eq.(\ref{fermion_mode_expn})---which constitutes the only support of
the Higgs.  The existence of non-diagonal (in the level space) Yukawa
as well as gauge couplings, both resulting from the localization of
the Higgs on co-dimension 2 branes, has an immense bearing on the
stability of the Higgs potential.  Quantum corrections to the quartic
Higgs vertex now emanate from a plethora of diagrams, with the
negative contributions from the multitude of top (and top-cousin)
loops, each proportional to $\lambda_t^4$, thereby, quickly destabilizing the
potential. The larger multitude of diagrams, potentially, renders the
problem even worse than that within the mUED~\cite{Datta:2012db}.
Consequently, the cutoff $\Lambda$ needs to be relatively low.

Furthermore, this scenario leads to a computational problem in
  that the calculation of the gauge-boson wave functions is rendered
  very complicated. While a product such as $\delta(x_4) \,
  \delta(x_5)$ is best handled by going over to polar coordinates,
  this method fails here, not only on account of the lack of a
  spherical symmetry, but also as we have to deal with four such
  products. One could, still, adopt such a method in neighborhoods
  around each corner, and then sew them together. This, however, is
  not very illuminating. An added consequence is that the consequent changes
  in the gauge-boson wavefunctions (which, naturally, depend on the
  ratio of the symmetry-breaking-generated term and the compactification radii)
  lead to significant modifications of their couplings with the fermion
  zero-modes. 
  
  A way out of such problems would be to allow the Higgs to
propagate in the bulk or even just the branes containing the quarks.
This would imply that, as far as the tree-level Yukawa interactions
(in particular, the ones
involving the top-sector) are concerned, KK-number is now, rendered 
a good symmetry. 
These vertices being level-diagonal greatly reduces the number of
fermion (top) loops contributing to the Higgs quartic coupling,
postponing any instability to much higher energies\footnote{Note that
  the reduction in gauge loops is not as drastic. Moreover, with a
  Higgs tower being introduced now, a further source of stabilization
  emerges.}.


\subsection{Higgs on the quark branes}
\label{4branehiggs}
We discuss next the possibility that the Higgs are localized on the
two 4-branes at $x_5 = 0, \pi$. Apart from the fact that, in the
effective four-dimensional theory, there is now a tower of scalars
instead of just the one, there is a further important change.  The
equations of motion for the gauge bosons, now, include two
delta-functions, at $x_5 = 0, \pi$ respectively. While the solution
thereof is straightforward, a key point needs to be appreciated.
Although the scale of electroweak symmetry breaking is small compared
to the Kaluza-Klein scale, the former cannot be treated as a simple
perturbation as the solution space in the presence of a delta-function
potential is different from that without. In particular, the
characteristics of the gauge boson ground state changes radically,
thereby leading to potentially large phenomenological changes.

The problem can be ameliorated if, instead of infinitesimally thin
branes (as we have been assuming so far), we consider, instead, fat
branes. This would lead to, amongst other changes, the tempering of
the delta-function, and, as can be ascertained without much effort,
the ground state would receive corrections of ${\cal O}(m_W^2 R_q^2)$
without destroying any of the important properties.

To execute such a possibility, we need to trap matter at specific
locations in space. An exceedingly simple mechanism was discussed
in \cite{Rubakov:1983bb} and all we need is a confining potential such
that a threshold energy is required to escape from the potential
well. Such a potential, for example could be formed by kinks in a
scalar theory.

\subsubsection{Kinks in six dimensions}
Consider a real scalar 
field $\chi$ with a Lagrangian given by~\cite{coleman}
\beq
{\cal L} = \frac{1}{2} \, \partial^M \chi \,\partial_M \chi - V(\chi) \ ,
  \qquad V(\chi) = \frac{1}{2 \, m^2 \, a}\ \left[a \, \chi^2- m^4 \right]^2,
\eeq
where $a$ is a positive constant and
$M=0,\dots,5$. Thus, the self-coupling is $a / m^{2}$. Although ${\cal
L}$ is invariant under the $Z_2$ symmetry $\chi \to - \chi$, the two
degenerate minima, viz. $\chi = \pm m^2/\sqrt{a}$, evidently do not
respect this $Z_2$. As is well-known, if a potential admits degenerate
vacua, nontrivial time independent solutions to the equation of
motion exist. While many different and inequivalent kink solutions
are possible, it is the boundary conditions that
dictate the appropriate one. Concentrating on classical solutions
that are nontrivial only along the $x_5$ direction, we have 
\[
   r_\ell^{-1} \, \partial_5 \, \chi(x) = \left[ - 2 \, V(\chi) \right]^{1/2}
\]
leading to
\beq
\chi_{cl}(x_{5})=\pm\frac{m^2}{\sqrt{a}} \,
            \tanh \left(\frac{ m \, r_\ell \, x_{5}}{\sqrt{2}}\right).
\eeq
Henceforth, we refer to the positive (negative) sign as the 
kink (antikink) solutions. 
The energy of each is given by 
\beq
E_{\rm kink}= r_\ell \, \int dx_5 [\partial_{5}\chi_{cl}(x_{5})]^2
  =\int{d\chi} \, [-2V(\chi_{cl})]^{1/2}=\frac{\sqrt{2} \, m^{5}}{a}. 
\eeq
The modes about the kink solution can be obtained by 
effecting a perturbative expansion about it, namely
$\chi(x^{M})=\chi_{cl}(x_{5})+\tilde \chi(x^{M})$. Linearizing 
the 
equation of motion for $\tilde \chi(x^{M})$, we have 
\[
\left[ \partial_M\partial^M - m^2 +\frac{3\, a}{m^2} \, \chi^2_{cl} \right]
    \tilde \chi=0.
\]
If we want to interpret this in terms of five-dimensional modes, 
we must re-express $\tilde \chi$ as 
\beq
\tilde \chi(x^{M}) = \sum_i \chi_{i}(x^{\bar M}) \, \eta_{i}(x_{5}),
\eeq
where $\bar M = 0, \dots, 4$ and $\eta_{i}(x_{5})$ form an orthonormal basis. 
The equation of motion, then, simplifies to 
\[
\eta_i(x_5) \, \partial_{\bar M}\partial^{\bar M} \chi_i( x^{\bar M})
    - \chi_i( x^{\bar M}) \, 
      \left[r_\ell^{-2} \, \partial_5^2 + m^2 - \, \frac{3\, a}{m^2} \, \chi^2_{cl} 
        \right] \eta_i(x_5) =0,
\]
where $\bar{M}=0,1,2,3,4$, and $\eta_{i}(w) $ form orthonormal
basis. Clearly, $\eta_i(x_5)$ has to be an eigenfunction of the
differential operator contained in the brackets. This is more
conveniently expressed in terms of a rescaled dimensionless variable
$z \equiv m \, r_\ell \, x_5 / \sqrt{2}$, to yield
\begin{equation}
\label{eq:kinkself}
\left[-\partial_z^2 -2+6 \, \tanh^2(z) \right]
 \eta_{i}(z)=\frac{2 \, \omega_{i}}{m^2} \, \eta_{i}
\end{equation}
\begin{figure}
\begin{center}
\includegraphics[scale=0.75]{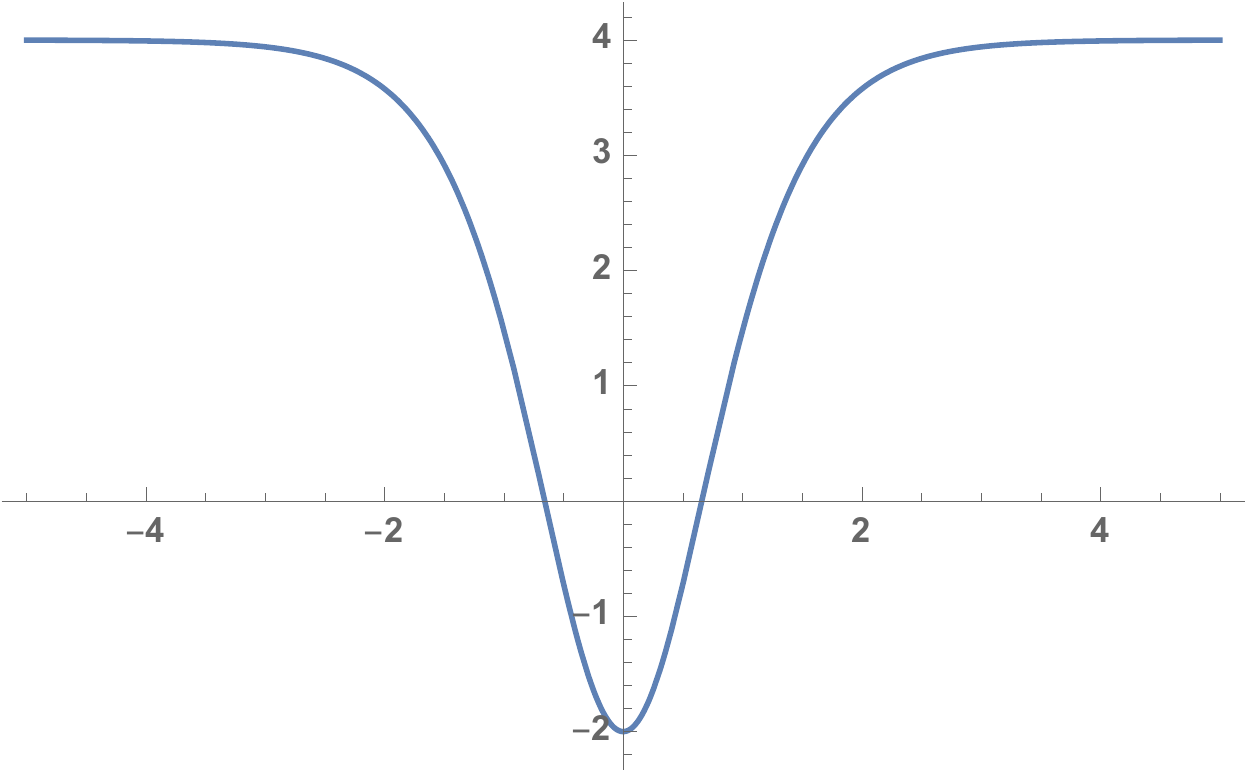}
\end{center}
\caption{potential due to scalar field symmetry breaking 
    to the equation $6\tanh[z]^2-2$.}
   \label{fig:potential}
\end{figure}
This is but a one-dimensional Schr\"odinger equation for a potential
as depicted in Fig. \ref{fig:potential}. Indeed, this is a particular
example of a general class of potentials
$U_\ell(z) \equiv \ell \,(\ell + 1) \,\tanh^2 z - 2$.  The problem is
well studied \cite{DamienP.George:2009crn} and the spectrum contains
exactly $\ell$ discrete states with a continuum beyond.

 \subsubsection{Higgs localization} 

The simplest lagrangian for the doublet Higgs $H$ including a
 $Z_2$--invariant interaction with the bulk kink field $\chi$ can be
 written as\footnote{The usual Higgs quartic term has been
 omitted as it is not germane to the issue.}
\beq
\barr{rcl}
{\cal L}_{scalar} &=& \dis \frac{1}{2}(\partial_{M}\chi) \, (\partial^M \chi)
                          - V(\chi) 
                 + \frac{1}{2} (D^M H) (D_M H^\dagger) 
                 - \, \frac{1}{2} M_H^2 (H^{\dagger}H)
\\[2ex]
&+ & \dis 
\frac{a}{2 m} g_{H \chi} (g_{H \chi}+1)\chi^2 |H|^2 
\earr
\label{lagrangian_scalar_flat}
\eeq
where $m$ is
the mass of the kink field.
The equation of motion for the free Higgs field, in the $\chi_{\rm cl}$ background, is then
\[
\Big(-\partial_{\bar{M}} \partial^{\bar{M}} -r_\ell^{-2}\partial_5^2 + \frac{a}{m} 
g_{H \chi} (g_{H \chi}+1)\chi_{\rm cl}^2 + M_H^2 \Big)H = 0 \ .
\]
 Writing the Higgs field as
\beq
H(x_\mu,x_4,x_5)= \sum_p H^{(p)}(x_\mu,x_4)f_p(x_5)
\eeq
where $f_p(x_5)$ are a set of orthonormal functions, 
 we get,
\beq
\Big(-M_p^2 -r_\ell^{-2} \partial_5^2 + \frac{1}{m}a g_{H \chi} (g_{H \chi}+1)\chi_{\rm cl}^2 + M_H^2 \Big)f_p = 0 \ ,
\label{eq:higgseom}
\eeq
where we consider $H^{(p)}$ to be solutions of $\partial_{\bar
 M} \partial^{\bar M} H^{(p)} = M_p^2 H^{(p)}$.  Taking the classical
 configuration for the kink field, $\chi_{cl}
 =\sqrt{m^3/a} \, \tanh(m \, r_\ell \, x_5/\sqrt{2})$,
 the solution to the first bound state of the above equation with mass
 $M_0^2= M_H^2 + \frac{1}{2}m^2 g_{H \chi}$ is found to be
\beq
f_0 = A_0 \cosh^{-g_{H\chi}}\left(\frac{m \, r_\ell \, x_5}{\sqrt{2}}\right)
\label{eq:tripletlocalize}
\eeq
For $g_{H \chi} = 1$, there exists only one bound state and the solution becomes
\[
f_0 = {\rm sech}\left(\frac{m \, r_\ell \, x_5}{\sqrt{2}}\right)
\]
and could be considered as Higgs for $ M_H^2 +
\frac{1}{2}m^2 g_{H \chi} = (125 \gev)^2$.

Once the geometry is compactified, we need a system with kink and an
anti-kink forming the domain walls at the boundaries of the orbifold.
This leads to the symmetric space we have along the $x_5$ direction. 
Since the Higgs Lagrangian is symmetric with respect to the kink
and anti-kink, the localization of Higgs would occur in the same way
on both the domain walls. Hence the bound state of Higgs would
have the wave
function
\[
f_0 = \frac{1}{2}\left[ 
{\rm sech}\left(\frac{ m \, r_\ell \, x_5}{\sqrt{2}}\right) 
+ {\rm sech}\left(\frac{ m \, r_\ell \, (\pi-x_5)}{\sqrt{2}}\right)
\right].
\]
Taking $m \rightarrow \infty$ we get back the delta function
localization of Higgs field. Analogous mechanisms exist for fermion 
 localization. We, however, will not dwell further on this mechanism as the 
 phenomenological ramifications, in some sense, lie in between the two other choices of 
 electroweak symmetry breaking.


\subsection{Bulk Higgs}
Allowing the Higgs field to
propagate in the entire bulk, while leading to an additional profusion 
of fields in the four-dimensional limit (in the form of 
a tower of towers of scalars), is actually the simplest, of the 
three cases, to handle. The Lagrangian for this case is given by
\begin{equation}
S=\int d^{4}x\intey\intez \, \sqrt{g} \left[(D_M H)^{\dagger}(D^M H) 
  + V(H^\dagger H)\right]
   \label{BulkHiggs_lagr}
\end{equation}
where 
\[
D_M = \partial_M - i \, g_1 \, B_M - i \, g_2
  \, W_M^a T^a
\]
and $V(H^\dagger H) =
 - \mu^2 \, H^\dagger H + \lambda \, (H^\dagger H)^2$. The mode expansion is straightforward, namely, 
\beq
\barr{rcl}
\pi \, \sqrt{R_q \, r_\ell}\, H(x^{\mu},x_{4},x_{5})& = & \dis  
H^{(0,0)}(x^{\mu}) 
   + 2 \, \sum_{j,k > 0} H^{(j,k)}(x^{\mu}) \;
          \cos\left(j x_{4}\right) \cos\left(k x_{5}\right)  
\\
         & + & \dis \sqrt{2} \sum_{k> 0} H^{(0,k)}(x^{\mu})\;
          \cos\left(k x_{5} \right) 
   + \sqrt{2} \, \sum_{j>0} H^{(j,0)}(x^{\mu})\;
          \cos\left(j x_{4} \right)
\earr
\eeq
Only the $H^{(0,0)}$ Higgs doublet develops a vacuum expectation value, namely $H^{(0,0)}= 
(0, v/\sqrt{2})$ and the zero modes of both the electroweak gauge bosons and the Higgs
boson acquire masses just as in the SM.
The higher modes (including the
charged higgs and pseudoscalar states) too mix within a given level, 
but with far smaller angles, owing to the hierarchy between the electroweak 
scale and the compactification scales. The mechanism
is similar to that operative in the mUED case\cite{CorderoCid:2011ja}, albeit with additional complications.

The Yukawa lagrangian is given by
\beq
\barr{rcl}
S& =& \dis 
\int d^{4}x\intey\intez \, \sqrt{g} \, \Bigg\{ 
     \left[\lambda_u \, \bar{Q} \, u \, (i\sigma^2 H^*) + \lambda_d \, \bar{Q} \, d \, H\right] \left[ \frac{\delta(x_5) + \delta(x_5-\pi)}{2} \right] 
\\[2ex]
& & \dis \hspace*{10em} + 
 \lambda_e \, \bar{L} \, e \, H\,\delta(x_4) \Bigg\}
\ ,
\earr
\eeq
where, for the sake of simplicity, we have suppressed the generational indices. With
the fermions
acquiring masses via boundary localized terms, there would exist mixings between the wavefunctions.
However, this mixing is, understandably,
 small except for the case of the top 
  quark. We shall not dwell on this aspect any further.

\section{Quantum Corrections to the Spectrum}
    \label{sec:spectrum}
As we have seen in the preceding sections, with the quarks being
localized symmetrically on the pair of end-of-world 4-branes at
$x_5=0$ and $x_5=\pi $, there exists a $Z_2$ symmetry, namely $x_5 \to
\pi - x_5$. This is reflected by the corresponding wavefunctions being
either symmetric (even $p$) or antisymmetric (odd $p$) about $x_5 =
\pi / 2$. This $Z_2^\ell$ is analogous to the KK-parity in mUED, but
operative only on modes along the $x_5$-direction. On the other hand,
with the leptons being localized on the 4-brane at $x_4 = 0$ alone,
the corresponding symmetry in this direction is lost entirely. Thus,
while the lightest of the ${\cal F}^{(0,1)}$ particles (where ${\cal
  F}$ is an arbitrary field in the theory) is stable, this is not true
for the ${\cal F}^{(1,0)}$.

The identification of the lightest of the level-1 excitations in
either direction proceeds quite analogously to that in mUED. At the
tree-level, the level-$(0,1)$ leptons and the $B_\mu^{(0,1)}$
are nearly degenerate. In the other direction, the level-$(1,0)$
partners of the light quarks are nearly degenerate with
$B_\mu^{(1,0)}$ and $G_\mu^{(1,0)}$. However, quantum corrections lift
these degeneracies. The corrections turn out to be small and negative for the
$B_\mu$-cousins, and positive for the others. The extent of this
splitting grows with the cutoff scale $\Lambda$.

While, within mUED, it has been argued that the stability of
electroweak vacuum\cite{Bhattacharyya:2006ym} dictates $\Lambda \leq
4\,R^{-1}$, this constraint is not strictly applicable here. Vertices
like $\bar t^{(n,0)} \, t^{(m,0)} \, H$ introduce additional box
diagrams which tend to drive the effective Higgs self-coupling
negative, thereby tending to destabilize the vacuum. These are only
partially offset by the contributions from the $W^{(n,m)}$ (and
$Z^{(n,m)}$) loops. The problem, however, is ameliorated to a
great extent by allowing the Higgs to traverse the entire quark brane
(thereby eliminating $\bar t^{(n,0)} \, t^{(m,0)} \, H^{(0,0)}$
vertices for $n \neq m$) instead of localizing it to the corners. To be
conservative, we take heart from the
fact that $\Lambda \sim 3 \, \max(R_q^{-1}, r_\ell^{-1})$ is a choice
quite safe, irrespective of the realization of EWSB, from the viewpoint of Higgs stability, and we adopt
this in this paper. To ease comparison with the literature (pertaining
to mUED) we also demonstrate results for a case with a different
  choice as well\footnote{Note that the use of two different
    $\Lambda$'s, one for each direction, is bad in spirit, for the
  theory should have only a single cutoff. One could, instead, argue
  for the inclusion of more complicated threshold terms to compensate
  for the existence of two compactification radii.}, namely, $\Lambda = 20 \,
\max(R_q^{-1}, r_\ell^{-1})$.

\begin{table}[!t]
\begin{center}
  \begin{tabular}{|c||c|}
\hline    Masses & Quantum Correction \\
\hline
    $m_{Q}$ & $\dis M \, \left[ 1 + \left(3\, a_3+ \frac{27}{16}\, a_2 + \frac{1}{16}\, a_1\right)\mathbb{L}\right]$\\
    $m_{u}$ & $\dis M \, \left[1 + \left(3\, a_3+ \, a_1\right) \mathbb{L} \right]$\\
    $m_{d}$ & $\dis M \, \left[1 + \left(3\, a_3  + \frac{1}{4}\, a_1\right) \mathbb{L} \right]$\\
    $m_{L}$ & $\dis M \, \left[1 + \left(\frac{27}{16}\, a_2 + \frac{9}{16}\, a_1\right) \mathbb{L} \right]$\\
    $m_{e}$ & $\dis M \, \left[1 + \frac{9}{4}\, a_1\mathbb{L} \right]$\\
    $m^2_{B_{(0,n)}}$ & $\dis M^2 \, \left[1 - \frac{a_1}{6} \, \mathbb{L} 
             - \frac{17 \, a_1}{2 \, \pi^2} \, \zeta(3) \right]$\\
    $m^2_{W_{(0,n)}}$ & $\dis M^2 \, \left[1 + \frac{15}{2}\, a_2\mathbb{L} + \frac{a_2}{2 \, \pi^2} \, \zeta(3)  \right]$\\
    $m^2_{B_{(n,0)}}$ & $\dis M^2 \, \left[1 -\frac{a_1}{6}\, \mathbb{L} 
     - \frac{21 \, a_1}{2 \, \pi^2}  \, \zeta(3) -8\, a_1 \, \mathbb{L} + \frac{m_h^2}{2 M^2} \, a_2\,\mathbb{L}\right]$\\
    $m^2_{W_{(n,0)}}$ & $\dis M^2 \, \left[1 + \frac{15}{2}\, a_2\mathbb{L} 
  + \frac{a_2}{2 \, \pi^2} \, \zeta ( 3 ) -4\, a_2\,\mathbb{L} 
   + \frac{m_h^2}{2 \, M^2}\, a_2\,\mathbb{L}\right]$\\
    $m^2_{G}$ & $\dis M^2 \, \left[1 + \frac{23}{2}\, a_3\mathbb{L} - 
    \frac{3 \, a_3}{2 \, \pi^2} \, \zeta(3)  \, \right]$\\
    $m^2_{H_{(0,n)}}$ & $\dis M^2 \, \left[1 + \frac{3}{4}\, a_1\mathbb{L} + 
    \frac{3}{2}\, a_2\mathbb{L} - \frac{\lambda}{16\pi^2}\mathbb{L}  \, \right]$\\
    $m^2_{H_{(n,0)}}$ & $\dis M^2 \, \left[1 + \frac{3}{4}\, a_1\mathbb{L} + 
    \frac{3}{2}\, a_2\mathbb{L} - \frac{\lambda}{16\pi^2}\mathbb{L}  \, \right]$\\[0.5ex]
\hline
    \end{tabular}
\caption{\em The one-loop corrections to the the masses of the KK excitations. 
In each case, $M$ refers to the corresponding tree-level masses.}
\end{center}

\label{tab:masscorr}
\end{table}

The mass corrections can be separated into two primary classes, namely
those due to bulk corrections and that due to the orbifolding. Though the 
calculations are quite straightforward~\cite{Cheng:2002iz,Freitas:2017afm}\footnote{We 
have also checked our analysis with the improved corrections given in \cite{Freitas:2017afm}. 
These modify the KK mass spectrum atmost by 5\%, majorly affecting the strong sector.  
As a result, the branching fraction of the resonance particles decreases by 2\% 
and hence, there is overall 1\% modification to the results given in the present case.}, care must be
taken of the fact that, in the present case, additional
contributions accrue on account of non-diagonal couplings. It is useful
to define 
\beq a_i \equiv \frac{g_i^2}{16 \, \pi^2} \ , \quad
\mathbb{L} \equiv \ln \frac{\Lambda^2}{\mu^2} \ , 
\eeq 
where $g_i$ are the (4-dimensional) gauge coupling
constants and $\mu$ is the renormalization scale. In terms of these,
the masses are as given in Table \ref{tab:masscorr}.  The terms
proportional to the Riemann $\zeta(3)$ function denote the correction
due to the orbifolding, while the rest are due to the various
loops. Of particular interest are the terms $8 \, a_1 \, \mathbb{L}$
($4 \, a_2 \, \mathbb{L}$) pertaining to $B_\mu^{(n,0)}$ ($W_\mu{a \, (n,0)}$).
Appearing on account of the non-diagonal coupling of these bosons to
lepton-pairs (originating, in turn, due to the broken $Z_2^q$), these
terms have no counterpart in mUED scenarios. These corrections are
quite significant (and, indeed are the dominant ones for $B_\mu^{(n,0)}$)
leading to enhanced mass-splitting between particles of the same
order.

The hypercharge-boson excited states $B_\mu^{(1,0)}$ and
$B_\mu^{(0,1)}$ are, thus, the lightest excitations in the respective
directions.  The latter, being stable, is the DM candidate, while the
former decays promptly and, predominantly, to the SM leptons. It is
interesting to note that, in the event of $r_\ell \approx R_q$, the DM
candidate is actually the heavier of the two.

\section{The Dark Side}
   \label{sec:dark_constraints}

\subsection{The Relic Density}

 Given the smallness of the mass splittings, in the early universe,
 the DM particle and the next-to-lightest KK particles would have
 decoupled around the same epoch. This can affect the relic abundance
 of DM in three ways.  Before we list these, though, it should be
 pointed out that, contrary to mUED-like scenarios, not all
 KK-excitations of similar masses behave similarly. While the
 excitations along the leptonic ($x_5$) direction behave analogously
 to the NLKPs of mUED-like scenarios, it is the next-to-lightest lepton-direction excitations
 (NLLE) that are germane to the issue with the rest of the NLKPs
 (relevant only if $R_q \approx r_\ell$) playing a subservient
 role. With this understanding,

\begin{itemize}
\item NLLEs, after decoupling from the thermal bath, would decay to the 
   lightest KK excitation {\em i.e.}, the DM, thereby increasing 
     the latter's number density.
 \item The NLLEs would also have been
   interacting with the other SM particles before they decoupled, to
   replenish the DM and can keep the DM in equilibrium a little
   longer, thereby diluting their number density.
\item Similarly, the NLLEs could also co-annihilate with the DM---to a pair of 
  SM particles---again maintaining it (and themselves) in equilibrium longer. 
\end{itemize}

The net effect would be determined by a complicated interplay of all
such effects. A key issue is whether the NLLE decouples from the SM
sector at or before the same epoch as the DM, or significantly
later. In the latter case, the number density of the NLLE at the epoch
of its own decoupling may be well below that of the DM, leading to
only low levels of replenishment.  In such a situation, it is often
the second effect above that wins the day. Note that this is quite in
contrast with the case of the mUED, where the inclusion of the
co-annihilation channels increases the relic abundance thereby
strengthening the upper bound on $R^{-1}$. This aspect would prove to
be crucial in the context of our model.

\begin{figure*}[htb]
\centering
 \subfigure[]{\label{fig:param1}\includegraphics[width=75mm]{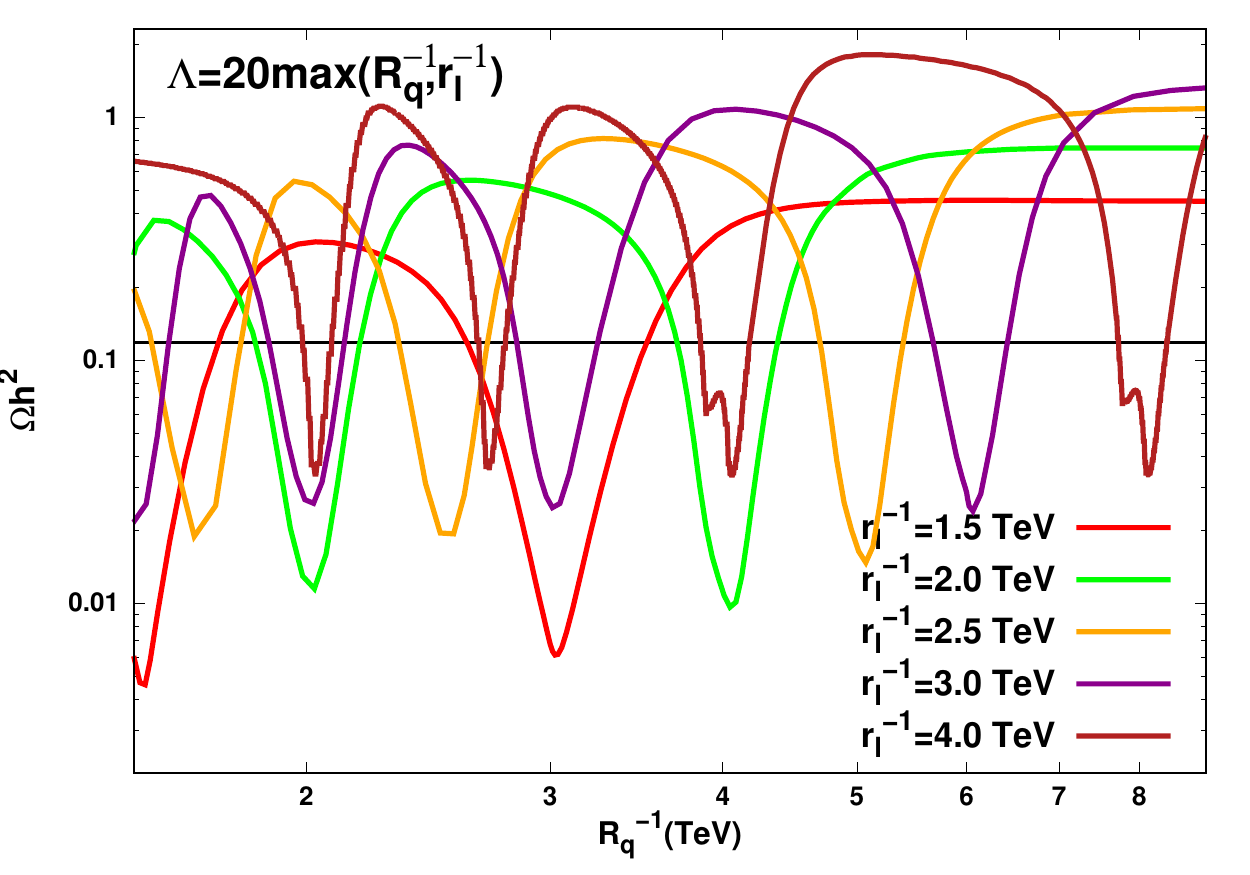}}
 \subfigure[]{\label{fig:param1b}\includegraphics[width=75mm]{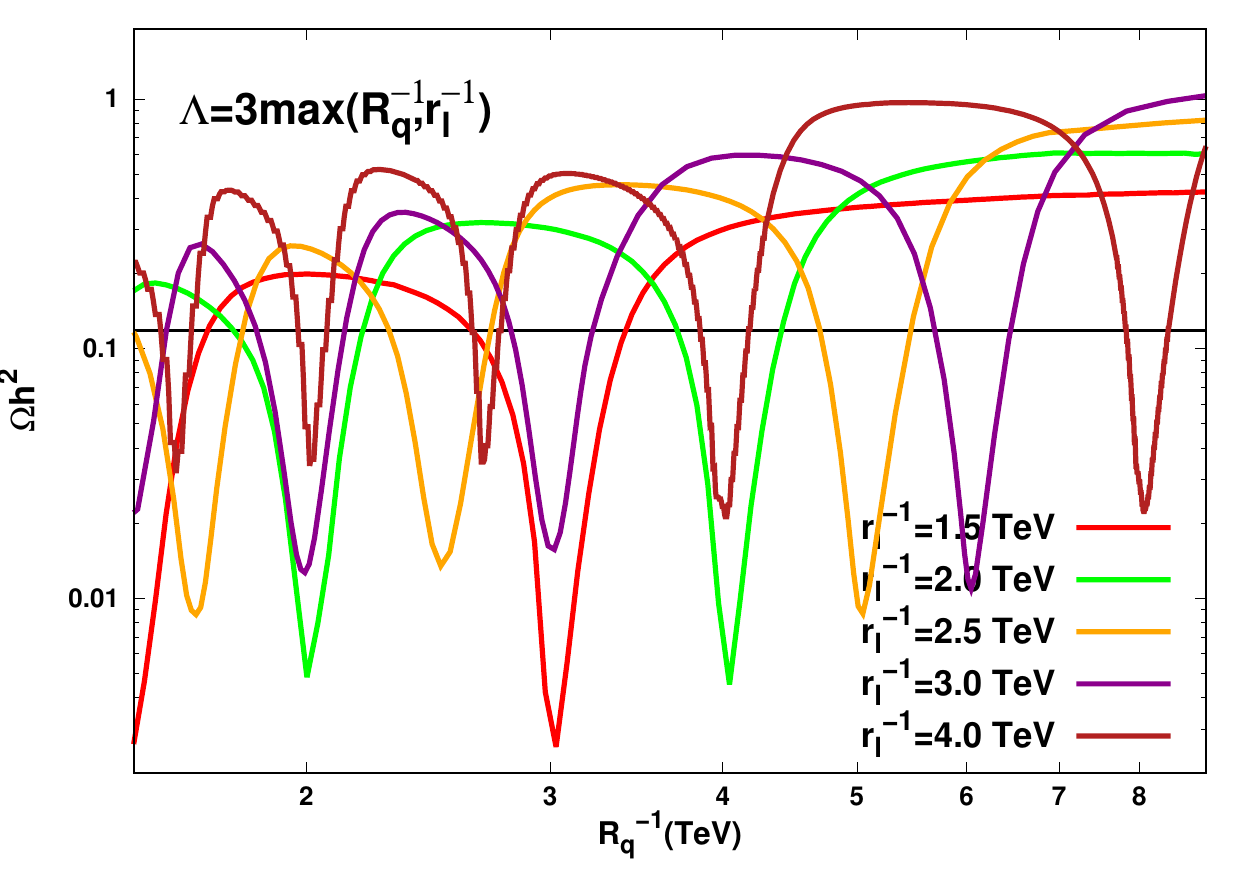}}
  \subfigure[]{\label{fig:param2}\includegraphics[width=75mm]{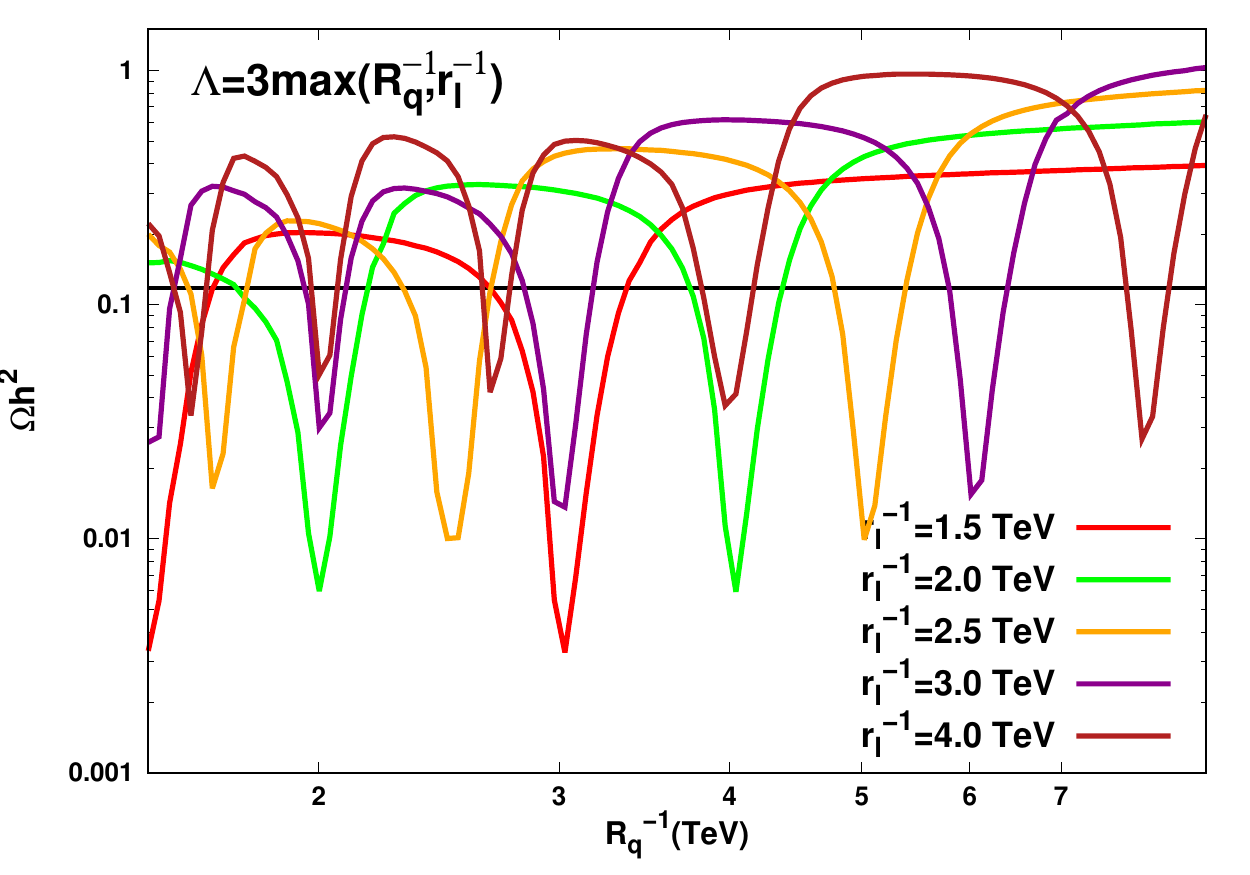}}
 \hspace{3cm}
\caption{\em $\Omega h^2$ as a function of $R_q^{-1}$ for different values of $r_\ell^{-1}$.
            (a) and (b) correspond to differing values of the cutoff $\Lambda$ for the case 
            when Higgs is localized at the 4 corners. (c) corresponds to the bulk Higgs scenario.}
\label{fig:param} 
\end{figure*}

To compute the relic density, we have implemented our model with the
interactions discussed in section \ref{sec:model} in
micrOMEGAs\cite{Belanger:2013oya} using
LanHEP\cite{Semenov:2014rea}. As a check, we have compared against the
CalcHEP model file discussed in Ref.~\cite{Belyaev:2012ai}.  Care must
be taken while calculating the relic density in micrOMEGAs. To produce
the plot in Fig.\ref{fig:param}, we have considered upto four KK
levels thereby requiring the modification of the
array size used in micrOMEGAs.
The resultant behavior of $\Omega h^2$ as a function of $R_q^{-1}$ is
depicted in Fig.\ref{fig:param}.  To understand the plots, several issues need to
be appreciated. We examine these in turn, concentrating first on
  the case of the Higgs located at the corners. With fewer particles in
  the play, this case is easier to understand, at least as far as the
  relic density is concerned.

\begin{itemize}

\item In mUED-like scenarios, there are a plethora of particles 
  nearly degenerate with the DM. While their co-annihilation with the 
  DM serves to drive down the latter's relic density, this is more than 
  offset by the twin effects of $(a)$ these interacting with the SM 
  particles prior to decoupling so as to replenish the DM and keep it 
  in equilibrium for a longer while, and $(b)$ once decoupled, these 
  decay into the DM, thereby enhancing the latter's relic density. Overall, 
  the existence of these particles serve to increase $\Omega_{DM} \, h^2$.

  In the present context, the excitations in the hadronic
    direction ($x_4$) play essentially no role in the aforementioned
    processes. Thus, one would expect the extent of the enhancement to
    be smaller.

\item A further crucial difference emanates from our having confined
  the Higgs field to the corners of the brane-box. Consequently, there
  are no Higgs KK-excitations. Within the mUED, the second-level
  excitations appear as $s$-channel propagators in processes such as
  $B_\mu^{(1)} B_\mu^{(1)} \to H^{0(2)} \to t^{(0)} \bar t^{(0)}$ and
  $B_\mu^{(1)} H^{+(1)} \to H^{+(2)} \to t^{(0)} \bar b^{(0)}$, with
  the fermionic vertices being generated at one-loop order. The
  suppression due to the loop factor is offset by
  the fact of these processes occurring close to resonance. (Note that
  the reverse process is not nearly as efficient.)

  With the absence of the Higgs-excitations in our model, this means of 
  suppression of the relic density is no longer available.

\item Overall, then, one would imagine that the relic density should
  depend on the mass of the DM particle ($\sim r_\ell^{-1}$) in much
  the same way as in the mUED (i.e when Higgs-excitations are not
  included). This naive expectation does describe the situation well
  for $R_q^{-1} \gg r_\ell^{-1}$ (the right part in
  Fig.\ref{fig:param}).

  As a closer study of Fig.\ref{fig:param} reveals, in this limit, the
  relic density, as a function of the DM-mass, is slightly higher than
  in the mUED case. This is expected due to the absence of the
  Higgs-excitations in our model.  Consequently, somewhat lower values
  of the DM-mass are now consistent with the Planck results for
  $R_q^{-1} \gg r_\ell^{-1}$.

\item A further feature is that the relic density increases with $\Lambda$. 
  This dependence is more pronounced away from the resonance. 
  This owes itself to the fact that larger $\Lambda$ leads to heavier NLLE, 
  and has a positive impact on $\Omega_{DM} h^2$.

\item To understand the shape of the $\Omega_{DM} h^2$--plots
  away from the right-edge, we need to remind ourselves of $B_\mu^{(n,0)}$
  (the excitations along the $x_4$--direction). Since these couple to
  all fermion pairs (including excitations), they mediate
  (unsuppressed) interactions between them. For $R_q^{-1} \approx 2 \,
  r_\ell^{-1}$, the $s$-channel diagram mediated by $B_\mu^{(1,0)}$ would
  be close to resonance, leading to highly enhanced
  cross-sections. Consequently, the NLLEs would remain in equilibrium
  (with the SM sector) until a later era. This reduces their number
  densities at the epochs of their own decoupling and, thereby, suppresses the
  replenishment of the DM number density through their decays. In
  addition, this late decoupling allows the co-annihilation processes
  to occur for longer time, further suppressing $\Omega_{DM} h^2$.

\item The suppression (in the relic density) discussed above is
  caused not by the $B_\mu^{(1,0)}$ alone, but by all $B_\mu^{(n,0)}$,
  with each coming into prominence when $2 \, r_\ell^{-1} \approx
  n \, R_q^{-1}$ (see Fig.\ref{fig:param}).  Numerically, even more
  important (on account of the gauge coupling $g_2$ being larger than
  $g_1$) are the roles of the $W_\mu^{a \, (n,0)}$. With these being
  close in mass with the $B_\mu^{(n,0)}$, the individual peaks cannot
  be distinguished in the plots.

  The shape of the individual dips is largely governed by two factors.
  The width of the gauge boson excitations is the primary one The
  slightly asymmetric nature is caused by the interplay of the
  cross-sectional behavior and dependence of the multitude of
  particle fluxes on the mass scale.

\end{itemize}

At this stage, we turn to the phenomenologically more interesting case
of the bulk Higgs. With this differing from the earlier case only in
the presence of additional scalars, one would expect that much of the
features described above would survive especially since the Higgs
excitations have suppressed (Yukawa) interactions with most of the
NLLEs. That this is indeed the case is borne out by
Fig.\ref{fig:param2}. There are some significant differences
though, especially away from the dips associated with $2 \,
r_\ell^{-1} \approx n \, R_q^{-1}$. These deviations owe themselves to
the presence of the Higgs KK-excitations. For example, we now have
$s$-channel processes\footnote{While many amplitudes receive
    contributions from the Higgs excitations, these are some of the
    dominant ones.} such as $B^{(0,1)}_\mu B^{(0,1)}_\mu\rightarrow
H^{(0,2)}\rightarrow t\bar{t}$ for the DM as also for the NLLE, such
as $W^{a(0,1)}_\mu W^{a(0,1)}_\mu\rightarrow H^{(0,2)}\rightarrow
t\bar{t}$. With these amplitudes being tree-level, unlike in the case
of the mUED, they could be expected to play a much more decisive role
here. And, indeed this is so, as the suppression in the relic density
away from the dips show. Near the dips, though, $B^{(0,1)}_\mu$
self-annihilation is not a very important issue. Instead, the $s$-channel 
(co-)annihilation diagrams mediated by the $B^{(n,0)}_\mu$ and $W^{a(n,0)}_\mu$
involve many more of the NLLEs, and, thus, play a much greater role in 
suppressing the relic density (see discussion above). Consequently, virtually 
no change is seen close to these dips. Having understood the small difference between 
Fig.\ref{fig:param1b} and Fig.\ref{fig:param2} in terms of the role played by the 
Higgs excitations, it is now easy to divine the outcome for a Higgs that is confined to 
the quark 4-branes. With this case being associated with a single tower of Higgs (rather than 
a tower of towers), one expects results that are in between the two cases. Indeed 
this is so, with the 4-brane results actually been close to the corner-localized case as the 
$H^{(0,2)}$ is entirely missing now. On the scale of the figures, the plots are virtually 
indistinguishable from each other.

It should be appreciated that, so far, the exploration of the parameter
space has paid no heed to other observables, such
as those at dedicated DM experiments or collider constraints. We turn
our attention to these next.

\subsection{Direct and Indirect search experiments}

Direct detection experiments have, traditionally, depended upon the DM
particle scattering (both elastic and inelastic) off nuclei.  In the
present context, the only tree-order diagram contributing to DM-quark
interactions is that mediated by the Higgs. This, naturally, is
suppressed by the size of the Yukawa coupling and is too small to be
of any consequence in the current experiments. It should also be noted
that a pair of $B_\mu^{(0,1)}$ cannot annihilate through a $s$-channel 
photon. The analogous contribution for $Z$ mediation, while not vanishing
identically, is again highly suppressed.

Some of the currently operating direct detection experiments are also
sensitive to DM-electron interactions. The sensitivity to the
effective coupling strength is lower, though (as compared to the
DM-nucleon interaction). In the present context, such scattering can
take place through $s$- and $t$-channel exchanges of the electron
excited states (namely, $e^{(0,1)}$ and $E^{(0,1)}$). Naively, it
might seem that a resonance is possible. However, the DM has very
little kinetic energy, and the electron too is not only
non-relativistic, but bound too.  Consequently, the cross sections are
too small to be of any interest.

\begin{figure*}[!h]
\centering
\includegraphics[width=0.5\textwidth]{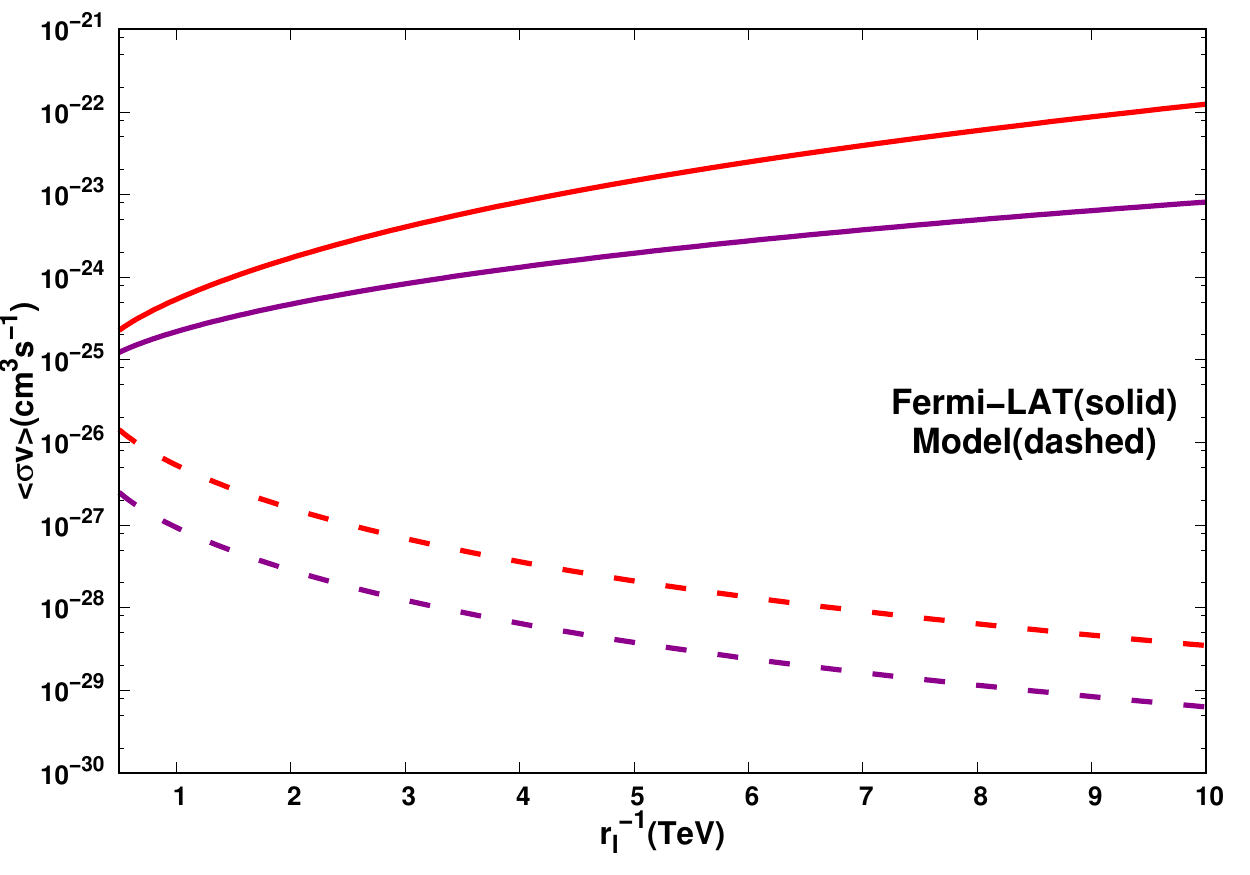}
\\\hspace{3cm}
        \caption{\em Thermal-averaged cross sections (dashed lines)
            for $B_\mu^{(0,1)} B_\mu^{(0,1)} \to W^+W^-$ (blue) and 
              $B_\mu^{(0,1)} B_\mu^{(0,1)} \to \tau^+\tau^-$ (red) as a function of the
            DM mass for $\Lambda = 3 \, {\rm max}(R_q^{-1}, r_\ell^{-1})$. 
            Also shown are the median upper limits on the DM
            annihilation cross-section as derived from a combined
            analysis of the Fermi-LAT data
            \cite{Ackermann:2015zua}.}
\label{fig:fermi}
\end{figure*}

Indirect detection proceeds through the annihilation of a DM-pair into
SM particles, which are then detected (typically, by satellite-based
detectors) either directly or through their cascades. In the present
case, a $B_\mu^{(0,1)}$-pair can annihilate into either a lepton-pair
($t$- and $u$-channel $e^{(0,1)}$ or $E^{(0,1)}$ exchanges) or to $W^+
W^- / ZZ / t \bar t / HH$ (all through a $s$-channel Higgs exchange).
This would be manifested in terms of both prompt and secondary continuum
emissions. The thermal-averaged annihilation cross sections for these
final states (as displayed in Fig.\ref{fig:fermi}) are, however, several orders
of magnitude below the most restrictive limits from
Fermi-LAT\cite{Ackermann:2015zua}.

\section{LHC Signatures}
   \label{sec:collider}
In the preceding section, we saw that this model admits much heavier
DM candidates (consistent with the relic abundance) than allowed
within the minimal-UED paradigm (whether 5-- or 6--dimensional). It
now behoves us to examine the collider phenomenology of the
same. Naively, a constriction of the moduli $(r_{\ell}, R_q)$ that
this model allows for would render the fields heavier, thereby
suppressing the production rates and easing the collider bounds as
compared to UED-like scenarios. On the other hand, the very structure
of the theory, namely that only one of the possibly two $Z_2$ parities
is conserved, brings forth new modes, both in the production arena as
well as in decays. To this end, we begin by reminding ourselves of some 
the particularly interesting couplings. 

\begin{figure*}[!t]
\begin{tabular}{ccc}
\includegraphics[width=5.4cm,height=5cm,bb= 0 0 560 505]{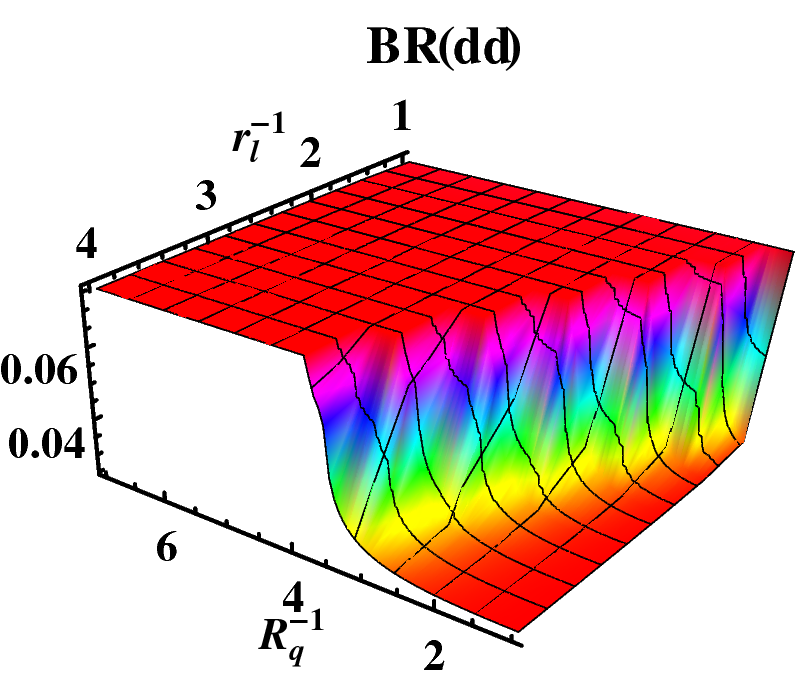} &
\includegraphics[width=5.4cm,height=5cm,bb= 0 0 560 505]{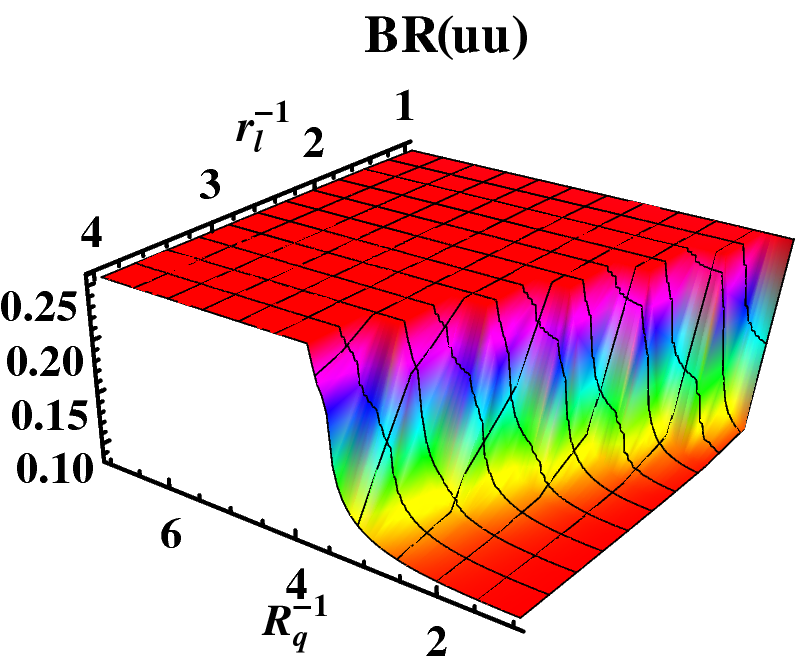} &
\includegraphics[width=5.4cm,height=5.2cm,bb= 0 0 560 505]{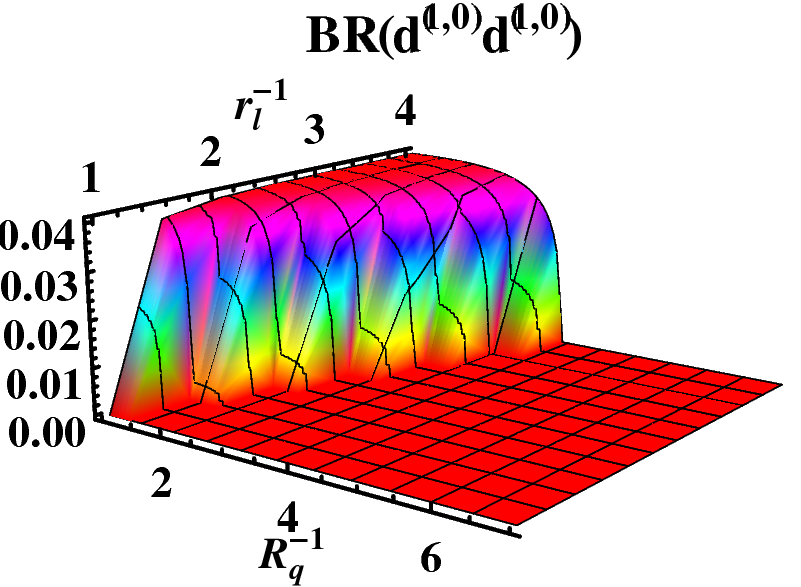} \\
\includegraphics[width=5.4cm,height=5.2cm,bb= 0 0 560 505]{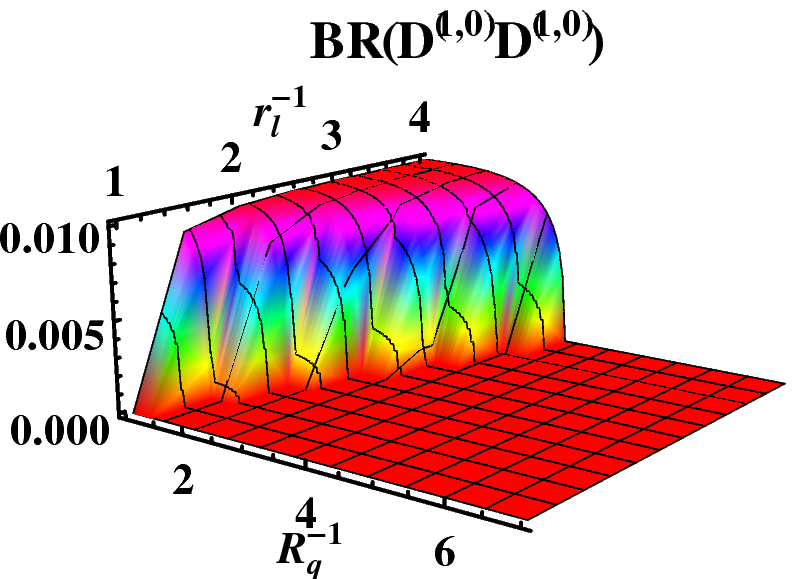} &
\includegraphics[width=5.4cm,height=5.2cm,bb= 0 0 560 505]{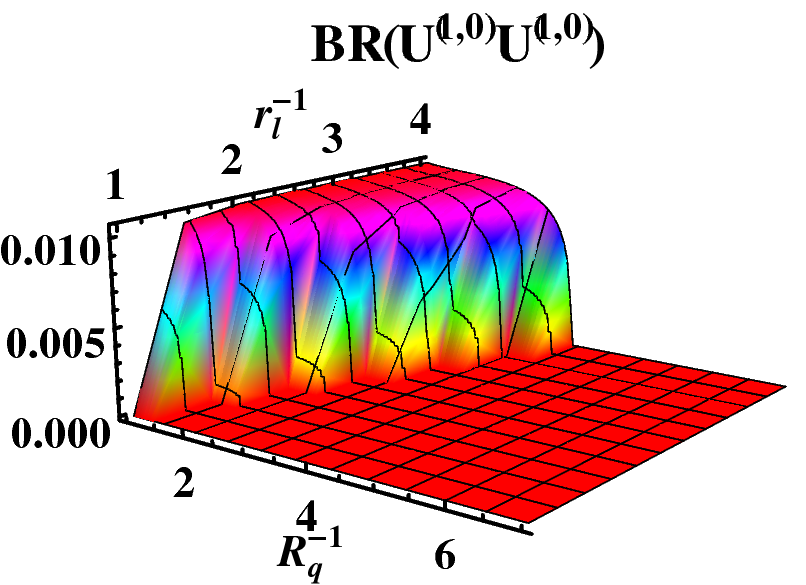} &
\includegraphics[width=5.4cm,height=5.2cm,bb= 0 0 560 505]{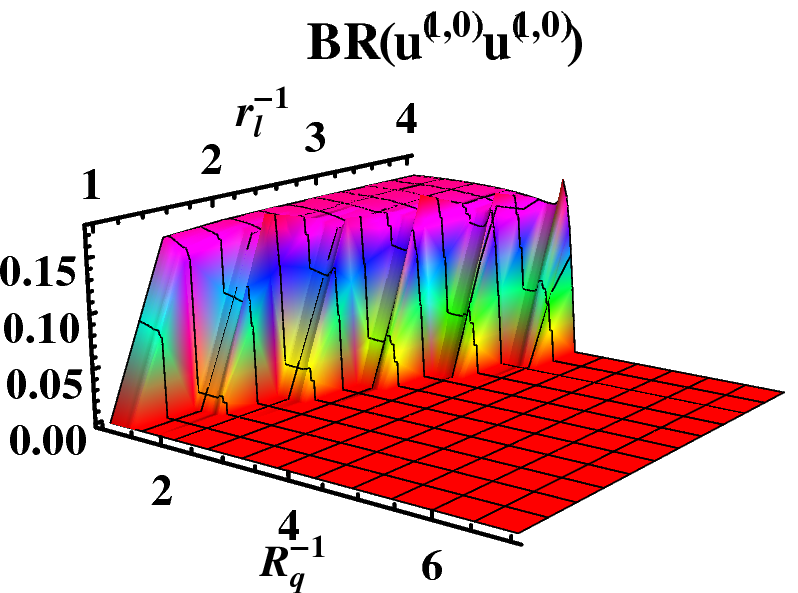} 
\end{tabular}
        \caption{\em{ Branching ratios of $B^{(0,2)}$ for down-type(d) and up-type quarks(u) as the final states. The quark singlets and doublets 
        are represented by $d$ and $D$ respectively. $\Lambda R=20$ has been assumed in the calculation.}}  
\label{fig:decay}
\end{figure*}

\begin{itemize}
\item Although $Z_2^{(q)}$ is not an exact symmetry, with the quark
  and gluon fields being confined to the 4-branes at $x_5 = 0,\pi$, 
  this symmetry is effectively an exact one in the context of
  the strong interactions. Thus, the first KK-state of the quarks
  ($q^{(1,0)}$) and gluons ($g^{(1,0)}$) can, essentially, be produced
  only in pairs.  The latter, being heavier, would decay predominantly
  into $q^{(1,0)} + \bar q^{(0,0)}$ pairs. The $q^{(1,0)}$, on the
  other hand, would decay into a SM-quark and the $B_\mu^{(1,0)}$ or a $W_\mu^{a \,
      (1,0)}$ with the branching fraction depending on the
  quark chirality. The bosons, in turn,
  would decay into leptons.

  The final state, from such production and decay channels, would,
  typically, comprise of (mostly soft) jets and charged leptons (or
  missing transverse momentum on account of neutrinos). The latter
  (MET) would be similar to the classic mUED signal and, hence,
  subject our model to the same constraints. For a mass of the parent
  particle similar to the mUED case, however, the fraction of events
  with a similar quantum of MET would be significantly
  smaller. While this suppression would be compensated by the presence
  of events with charged leptons, it is clear that the ensuing
  constraints would be similar to those applicable to the mUED
  case. Thus, a simple increase of the scale ($R_q^{-1} \gapp 2 \tev$)
  would maintain consistency with non-observation at the LHC.

\item An interesting alternative would be the resonant production $p p
  \to g^{(2,0)} + X$ starting with a $q \bar q$ pair (or, the
  analogous $q + g \to q^{(2,0)}$). Either of these vertices are
  loop-suppressed, leading to small cross sections. Once again,
  $g^{(2,0)} \to q^{(1,0)} + \bar q^{(1,0)}$ with the subsequent
  cascades as in the preceding cases. The signal too is similar, but
  of a markedly smaller size. It might
  seem that a invariant mass reconstruction (possible, in principle,
  for the $4 \ell + $ soft jets final state) would increase the signal
  significance. Given the smallness of the signal cross section, and
  the softness of the jets (leading to potential confusion with other
  sources such as pileups and multiple collisions), this is quite
  unlikely. A more definitive statement would require a full
  simulation beyond the scope of this paper.

\item More interestingly, a SM $\bar q \, q^{(')}$ pair has
  unsuppressed couplings with each\footnote{Similar is the story with
    $B_\mu^{(0,2n)}$ and $W_\mu{a \, (0,2n)}$ for all $n \in \mathbb{Z}^+$. The
    higher states, though, have nontrivial branching fractions into
    the quark KK-states, modes that are not necessarily
    available for $n = 1$.} of $B_\mu^{(0,2)}$ and $W_\mu^{a\, (0,2)}$. The
  latter, though, have only loop-suppressed couplings with a pair of
  SM leptons.  Thus, for $r_l^{-1} < R_q^{-1}$, the charged gauge
  bosons would manifest themselves as a leptophobic $W'$ while the two
  neutral ones would act like (nearly degenerate) pair of leptophobic
  $Z'$s.

The CMS collaboration has studied dijet final states in the quest for 
such resonances and the absence of any anomalies has led to constraints 
of the form $m_{W'} > 2.7 \tev$ and $m_{Z'} > 2.1 \tev$~\cite{Sirunyan:2016iap}. 
It should be noted that the inclusion of all the four electroweak $(0,2)$ 
KK-states would translate to a stronger bound. Furthermore, the aforementioned 
limits were obtained from the analysis of only $12.9\fb^{-1}$ of data
and assuming that no such anomaly would show up in the current round would 
push the bound up considerably. Thus, it might be safely assumed 
that $r_\ell^{-1} \lapp 1.7 \tev$ would be strongly disfavored. 

\item
For $r_\ell^{-1} > R_q^{-1}$, on the other hand, the situation is
altered quite dramatically. The $B_\mu^{(0,2)}$ and $W_\mu^{a\,
  (0,2)}$ can, now, decay to the first KK excitations of the quarks as
well. The branching ratio into SM quarks decreases while that into the
first KK quark states increases as $r_l^{-1}$ becomes progressively
greater than $R_q^{-1}$.  Furthermore, note that the coupling of the
$B_\mu^{(0,2)}$ to the singlet up-type quarks is larger than those to
the other quarks.  These features are reflected in the plots as
presented in Fig.~\ref{fig:decay}.

The level-1 quarks would further decay into a SM quark and a
$B_\mu^{(1,0)}$ or $W_\mu^{a\, (1,0)}$, with the latter cascading down
to leptons.  In the final analysis, such a chain 
would result in a final state comprised
of soft jets and leptons.  This signal is, understandably, more difficult to
analyze (as compared to the dijet signal discussed in the previous 
instant) but can still act as a complementary signal.

\item 
There are various
studies~\cite{Burdman:2006gy,Choudhury:2011jk,Burdman:2016njl} on
the collider phenomenology of 6D models like $T_2 /Z_2$ or $T_2
/Z_4$. However, the findings of the aforementioned studies
  cannot be directly translated to our case. With the mass spectrum
and couplings in the older models (studied
  there) being very different from the model under
consideration,  most of the signals and strategies
suggested in the earlier analyses are
not applicable here. The major difference is due to the fact that
$Z_2^{(q)}$ is not an exact symmetry.  As a result, the second KK
level gauge bosons (along quark brane) and quarks decay to SM leptons
at tree level in contrast to the cases considered in
Refs.\cite{Burdman:2006gy,Choudhury:2011jk,Burdman:2016njl} where
the corresponding decays only occur at the loop level.

For very similar reasons, even recent
  studies~\cite{Choudhury:2016tff,Deutschmann:2017bth,Beuria:2017jez}
  that seek to use experimental analyses in multiple channels
  (multijets, or dileptons with jets, each accompanied by missing
  transverse energy) also fail to apply. Rather, one needs to consider
  the event topologies described above.

\item As we shall see in the next section, low energy
  constraints, unless tamed by the introduction of compensatory fields,
  tend to push up $R_q^{-1}$ and, hence, the quark and gluon
  resonances to levels that are not easily accessible at the LHC. One
  would, then, have to concentrate on the KK-excitations of the
  leptons and electroweak bosons, and, consequently, devise algorithms
  more sensitive than those that have been deployed so far.

\end{itemize} 

\section{Low energy constraints}
\label{precision}
Stronger constraints emanate from low-energy physics, in particular,
the observables precisely measured at LEP and the SLD.

\subsection{Effective four fermi interaction} 

An important consequence of the fermions being brane-localized, while
the gauge bosons traverse the bulk, is that KK-number violating
gauge-fermion couplings appear. Their presence immediately results in
the alteration of the effective four-fermion interactions at
low-energies, the strength for which is usually parametrized in terms
of $G_f$.

In the present case, there exists a further subtlety. For the SM
leptons, only the $W^{a,\, (n, 0)}$ and the $B^{a,\, (n, 0)}$
contribute at the tree level. For the quarks, on the other hand, the
$Z_2$ symmetry ensures that the odd modes do not, and only the
$W^{a,\, (0, 2n)}$ and the $B^{a,\, (0, 2n)}$ contribute. Furthermore,
none of these KK-excitations couple to both quarks and leptons.

The best measurement of $G_f$ is given by $G_\mu$, the effective 
coupling strength for the muon decay. In the present case, this 
gets modified to
\beq
\barr{rcl}
G_\mu& =& \dis 
G_\mu^{(SM)}\left[1+ \sum_{n>0} 
      \left(\frac{g^{(n,0)}m_W}{g \, m_{W^{(n,0)}}}\right)^2\right]
\\
&\equiv&G_\mu^{(SM)} \, (1+V).
\earr
\eeq
Using, $g^{(n,0)}=\sqrt{2}\, g$ and simplifying the sum by neglecting the 
electroweak contribution to the masses in comparison to the 
KK-contribution, namely
\[
\sum_{n>0} \frac{1}{m_{W^{(n,0)}}^2} \gapp  R_q^2 \, \sum_{n>0} \frac{1}{n^2} 
\]
we have\footnote{This expression includes the entire tower. However, 
  a ultraviolet cutoff needs to be imposed for such theories (see the 
  discussion in the following section), and the ensuing contribution 
  is actually smaller, leading to a weaker constraint on $R_q^{-1}$.}
\[
V\approx 2 \, \zeta(2) \, m_W^2 \, R_q^2.
\]
Electroweak precision data constrains $V\leq0.0013$ implying
$R_q^{-1}\lapp 4 \tev$.

Had the four-quark effective operator been measured as accurately as
$G_\mu$, we would have been led to a bound on $r_\ell^{-1}$ that is a
factor of 2 weaker.  However, low energy data on this is not as
precise, with the inaccuracy exacerbated by the lack of a full
understanding of bound-state effects in hadrons. Consequently, the
corresponding bounds are actually much weaker and of little relevance
to us.

\subsection{Oblique Variables}
We present here the one-loop contributions to the electroweak
precision observables. We adopt the parametrization due to
Ref.\cite{Peskin:1991sw}, and omit the third parameter ($U$) as it is
of little relevance in the present case. The definitions, in terms of
the self-energies, are
\beq
\barr{rcl}
\dis \frac{\alpha}{4s_w^2\,c_w^2}\, S &=& \dis
\frac{\Pi_{ZZ}(m_Z^2)-\Pi_{ZZ}(0)}{m_Z^2}-
 \left.\frac{\partial\Pi_{\gamma\gamma}(p^2)}{\partial p^2}\right|_{p^2=0}-\frac{c_w^2-s_w^2}{c_w\,s_w}\left.\frac{\partial\Pi_{\gamma\,Z}(p^2)}
 {\partial p^2}\right|_{p^2=0}
\\[2ex]
 \alpha\, T& = & \dis 
\frac{\Pi_{WW}(0)}{m_W^2}-\frac{\Pi_{ZZ}(0)}{m_Z^2}-2c_w\,s_w\frac{\Pi_{Z\gamma}(0)}{m_W^2}
\earr
\eeq
Since the SM contributions are well known, we consider here only the 
deviations from the SM values, namely,
\beq
\mathcal{S} \equiv S - S_{SM} \ , \qquad
\mathcal{T} \equiv T - T_{SM} \ .
\eeq
In calculating the loops, we neglect further small corrections
(wherever applicable) wrought by the changes in the wavefunctions.
While the observables ${\cal S,T}$ are finite (and, independent of the
renormalization scale), the individual loops are divergent and are
calculated using dimensional regularization.  Using the expressions
derived in the Appendix, we list below the individual contributions
wrought by the various fields:
\begin{itemize}
\item \underline{Fermions}\\ 
On account of the large Yukawa coupling, the contributions due to the
top-partners far outweigh the rest and we have, for an individual 
excitation of mass $m_n$,
 \beq
 \mathcal{S}\approx \frac{m_t^2}{12\pi\,m_n^2} \ , \qquad 
\mathcal{T}\approx \frac{m_t^4}{8\,\pi\,s_w^2\,m_n^2\,m_W^2}.
\eeq

\item \underline{Gauge bosons}\\ 
These contributions are very small, in particular to ${\cal S}$.  For
the $(n,0)$ and $(0,n)$ states, we have the individual contributions
to be
\beq
\mathcal{S}\approx \frac{3m_Z^2s_w^4}{5\pi\,m_n^2}
\ , \qquad
\mathcal{T}\approx \frac{m_W^2}{\pi\,c_w^2\,m_n^2} \ ,
\eeq
whereas for the excitations with non-zero KK-numbers in both
directions, an additional factor of 1/2 appears for both 
${\cal S, T}$.

\item \underline{Higgs bosons}\\
If the Higgs is corner-located, there are no additional loops and the
only additional contribution accrues from the changes in the
wavefunctions.  For the 4-brane localized Higgs, on the other hand,
there is a single tower of scalars to be considered, in addition to
contributions from the changes in the gauge-boson wavefunctions as
alluded to in Sec.\label{4brane_higgs}. And, finally, for bulk higgs, 
corrections accrue only from the tower of towers (of scalars). The loop 
corrections, for $(n,0)$ and $(0,n)$ Higgs bosons, are
 \beq 
\mathcal{S} \approx \frac{m_h^2-3m_Z^2}{12\pi m_n^2} \ , 
\qquad
\mathcal{T} \approx - \, \frac{5m_H^2}{24\,\pi\,c_w^2\,m_n^2}
                   - \, \frac{7\,m_W^2}{24\,\pi\,c_w^2 m_n^2}
\eeq
whereas for those with non-zero KK-numbers in both
directions, an additional factor of 1/2 appears as in the 
gauge boson case. 
\end{itemize}

\begin{figure*}[htb]
\centering
\includegraphics[width=7cm,height=6.5cm]{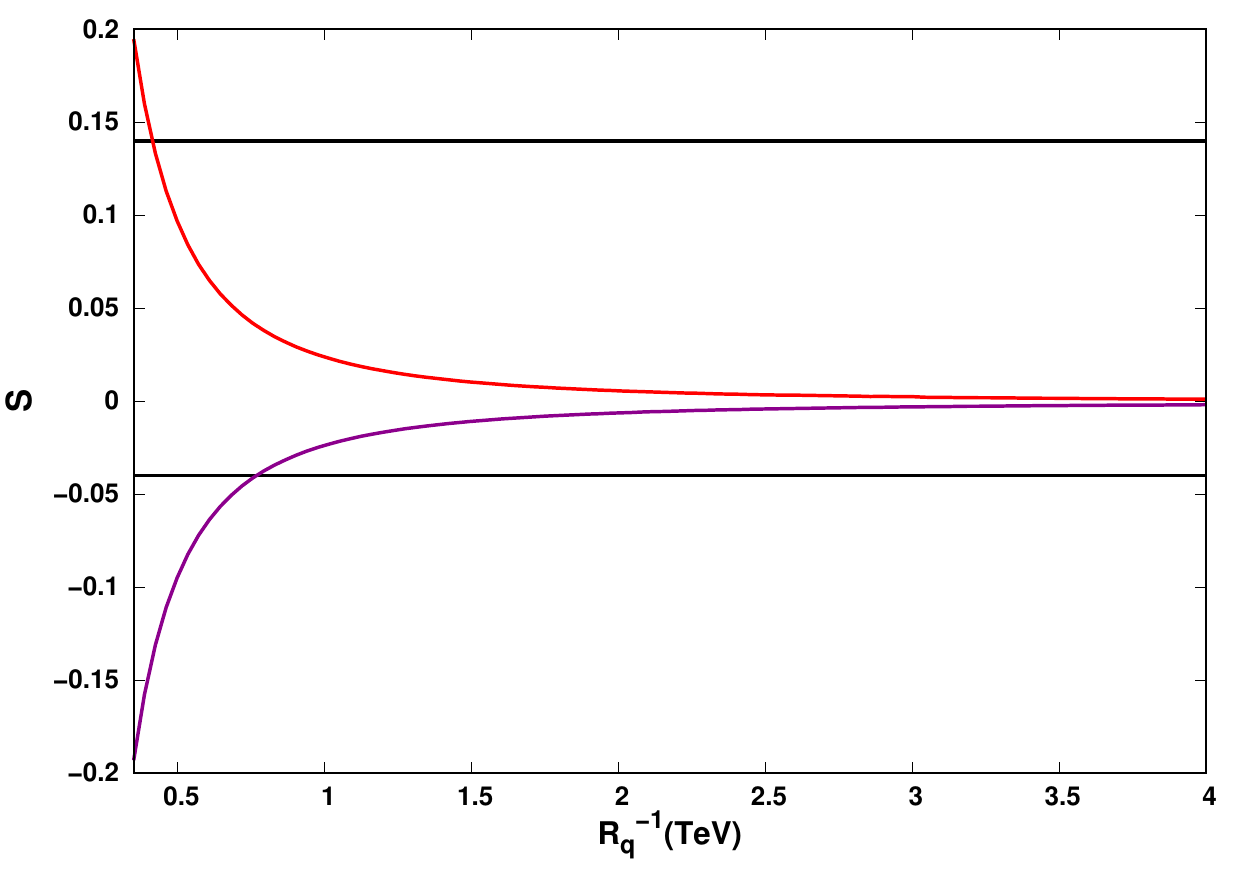} 
\includegraphics[width=7cm,height=6.5cm]{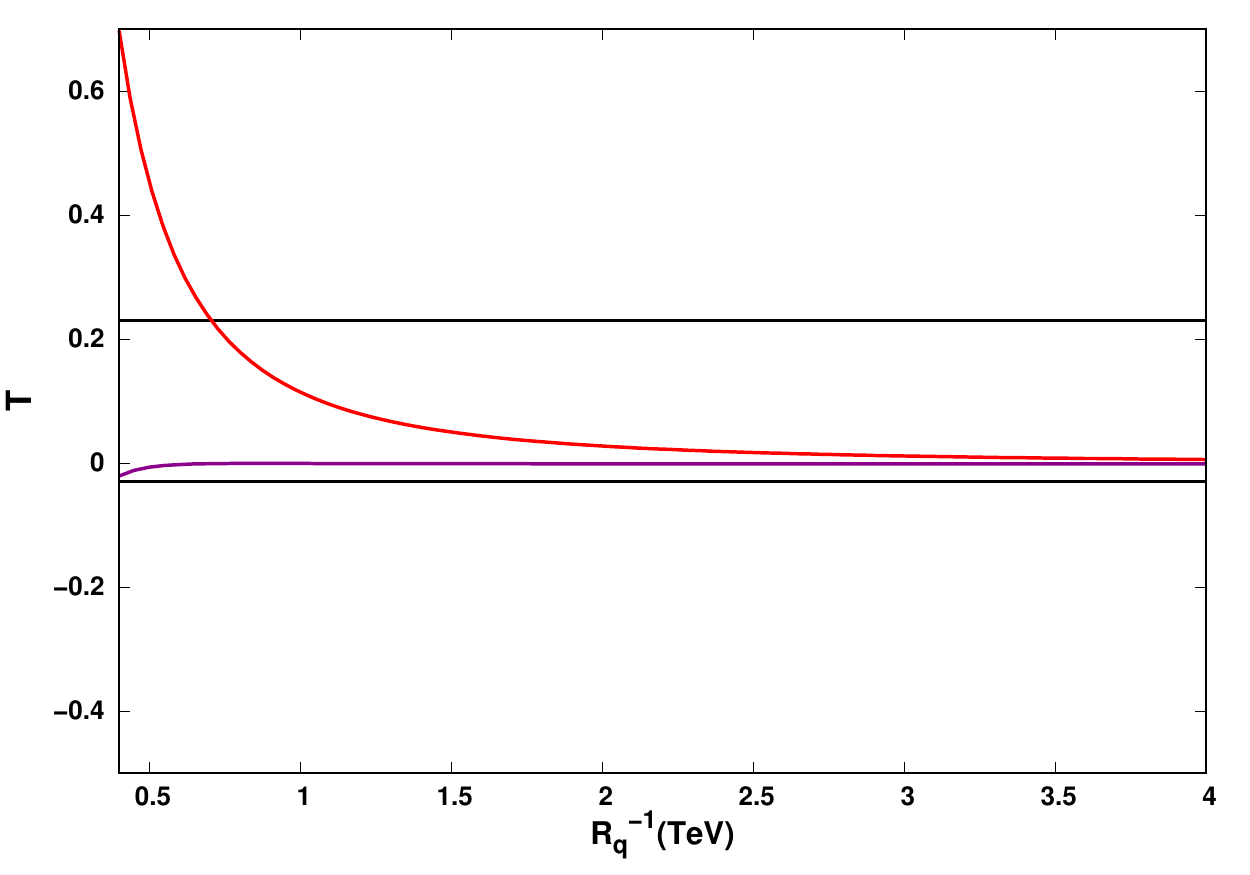} 
\hspace{3cm}
\caption{\em ${\cal S}$ vs $R_q^{-1}$ and ${\cal T}$ vs $R_q^{-1}$ obtained using 
$c_{1,2}=1$(red) and $c_{1,2}=-1$(blue). The horizontal bands correspond
to the 95\% C.L experimental constraints.
}
\label{fig:ST}
\end{figure*}
The individual contributions have, of course, to be summed over the
respective towers, and herein lies a problem. For a single tower, the
sum is a finite one. Exactly as in the preceding section, in the limit
of a vanishing electroweak scale---as compared to the compactification
scale---this can be expressed in terms of $\zeta(2)$.

Neglecting, for the time being, the six-dimensional nature of the
bosonic fields, the contribution from the top sector far overwhelms
the others. (Indeed, in this approximation, the contribution due to
Higgs and gauge bosons to T parameter is 5\%.).  Consequently, the
${\cal S,T}$ parameters are expected to primarily constrain
$R_q^{-1}$, with only minor sensitivity to $r_\ell^{-1}$. 

However, for a tower of towers (as is the case for a six dimensional field),
the sum is logarithmically divergent. Though the sensitivity to the
cut off is only a logarithmic one, we still cannot expect a reliable
estimate for the $S$ and $T$ parameters by only summing the KK
modes. The physics above cut-off ($\Lambda$) of the theory is
relevant, and the corresponding contribution to the parameters can be
roughly estimated using higher dimension custodial symmetry breaking
operators as discussed in Ref.\cite{Appelquist:2002wb}, leading to
 \beq
 \barr{rcl}
 T^{\text{UV}}&=&\dis c_1\frac{m_H^2}{4\Lambda^2\alpha(m_Z)}\\
 S^{\text{UV}}&=&\dis c_2\frac{2\pi v^2}{\Lambda^2}
 \earr
\label{eqn:ST}
  \eeq
Here, $c_{1,2}$ are constants giving the extent of custodial symmetry
breaking and deviation of coupling of $W^3$ and hypercharge $B$ gauge
field from the SM case respectively. Assuming maximal custodial symmetry violation, we
consider both $c_{1,2} = \pm 1$.

The values of $T^{\text{UV}}$ and $S^{\text{UV}}$ are dependent only
on the cut-off, which, in turn, is dependent on the number of
dimensions. As we have noted earlier, the stability of the electroweak
vacuum as well considerations of the naturalness of Higgs
mass~\cite{Appelquist:2002wb, Appelquist:2000nn} indicates that the
maximal cutoff is $\Lambda=3 \,R^{-1}$.  This is
the value of $\Lambda$ we assume in producing Fig.\ref{fig:ST}.  This,
in turn, implies that in calculating the loop contributions, only
fields with masses less than $\Lambda$ are to be included.
Interestingly, for the chosen values of $|c_{1,2}| = 1$, the contribution 
of the operators of eqn(\ref{eqn:ST}) to the ${\cal T}$-parameter is equal
to that of loop corrections due to quarks while in case of ${\cal S}$, it is 
entirely dominating.

As can be gleaned from Fig.\ref{fig:ST}, the constraints from 
the electroweak precision variables, resulting from the 
experimental measurements~\cite{Baak:2014ora}, which for ${\cal U} = 0$, read
\[
 {\cal S} =0.05\pm0.09 \ , \qquad 
{\cal T} =0.1\pm0.13 \qquad \mbox{(errors at 95\% C.L.)}, 
\]
are very weak indeed. Evidently, they pale in comparison to the restrictions imposed by 
the other observables.

\section{Summary}
\label{sec:concl}

Non-observation of any new state at the LHC has severely constrained
the parameter space for a host of well-motivated theories going beyond
the SM. In particular, the minimal UED suffers from the problem that
this non-observation militates strongly against the upper limit on the
compactification scale dictated by the observed DM relic density. The
ensuing tension can be relieved, to an extent, by invoking a
non-minimal theory with boundary-localized terms (respecting a $Z_2$
symmetry so that an explanation for the DM can be retained). However,
the continuing absence of any signals requires the size of such terms
(ostensibly, the result of quantum corrections) to become ever larger, 
thereby severely straining the entire paradigm. 

To mitigate the problems faced by the minimal UED, in this work, we
considered a quasi-universal six-dimensional theory, which naturally
inherits a much richer phenomenology compared to its five-dimensional
(or, even, usual six-dimensional) counterpart. Apart from the
interesting double-tower signatures, this scenario brings in many more
manifestations of possible BSM scenarios, due to the distinctively
different orbifolding of the space and field-localizations.

The scenario is quasi-universal in the sense that the quarks (and
gluons) and leptons are localized on orthogonal branes, while only the
electroweak gauge bosons traverse the entire bulk. This decouples the
masses of the quark and leptonic towers (by allowing for different
compactification radii), thereby enabling one to maintain the putative
DM at a suitably low mass, while rendering the quark and gluon
excitations relatively inaccessible at the LHC.

Naively duplicating the canonical mUED structure, however, is not
phenomenologically viable as this would have left behind a pair of
$Z_2$-symmetries, thereby leaving behind three DM fields, one each for
the lightest excitations in the $(1, 0)$, $(0,1)$ and $(1,1)$ sectors.
Two of these masses would be determined by the quark-excitation scale
and would run foul of the relic density measurements.

To this end, we propose that the quarks (and gluons) are localized on a pair
of parallel end-of-the-world 4-branes, while the leptonic fields are localized on a single such brane orthogonal to the other
two. The electroweak gauge bosons must, obviously, extend across the
entire bulk. For the Higgs, several alternatives are possibles, such
as localizing them to the corners of the brane-box, allowing them to be
localized on the boundary-branes or to propagate in the entire bulk.
Localization, whether at corners or on branes, leads to
  boundary localized mass terms that can significantly modify the
  gauge boson wavefunction and their coupling with SM fermions. Thus
  the simplest alternative, devoid of such issues, is to allow the Higgs
  field to traverse the entire bulk.

The hard breaking of one of the two $Z_2$'s engenders unsuppressed
tree-level couplings, amongst others, for certain level-1
KK-excitations with a pair of SM fields. This not only renders two of
the putative DM candidates unstable (leaving behind the
$B_\mu^{(0,1)}$ as the only cosmologically stable particle), but also
has immense consequences as far as LHC signals are concerned. The
prompt decay of all the excitations along the quark-direction into SM
particles severely depletes the missing transverse momentum signal, a
cornerstone of LHC search strategies. Instead, they are manifested in
terms of a dijet final state, or final states with leptons and soft
jets. As of the present instant, it is the former modes that present
the strongest constraints, with the negative searches for leptophobic
$W'/Z'$, together, requiring that a DM mass $\lapp 1.7 \tev$ would be
strongly disfavored. However, even stronger bounds are imposed
 by low energy experiments. The KK number
  violating $W^{a(n,0)}\bar{\ell}{\ell}$ couplings modify the
  effective four-lepton interactions at low energies and the high-precision
  measurement of $G_\mu$ imposes a constraint $R_q^{-1}\lapp 4\tev$. It
  should be realized, though, this bound is strictly applicable only when
  the contribution of the infinite tower is included. For any 
such theory with a finite cutoff (as it must have), the bound weakens to an 
appreciable degree. The corresponding bounds on $r_\ell^{-1}$ are weaker
by more than a factor of 2, not only on account of hadronic uncertainties,
but also because only half as many $W^{a(0,n)}$s (namely, the even modes alone)
mediate the four-quark interaction term. The bounds from electroweak 
precision variables ${\cal S}$ and ${\cal T}$ are much weaker in comparison 
to that from the measurement of $G_\mu$.

Even the (weaker) LHC bound is stronger than that imposed, by
considerations of the relic density, on the mUED candidate for the
DM. In other words, this seems to bely our stated objective of easing
the twin-constraints (LHC and DM). However, the aforementioned
breaking of one of the $Z_2$'s manifests in a nontrivial way in the
determination of the number densities. With the quark-direction
excitations of the electroweak gauge bosons now having unsuppressed
couplings with all the SM fermions, as well as their own
(unidirectional) excitations, these can mediate interactions between
the SM particles and the first excited states of the leptons (which
are close to the DM in mass).  As a consequence, these leptonic states
remain in equilibrium until a later epoch, thereby suppressing their
density at decoupling. This, in turn, translates to their decays---into
the SM leptons and the DM---no longer being effective means to replenish 
the relic density. Since this replenishment is actually the major contributor 
to the relic density in UED-like scenarios, this effect can suppress 
$\Omega_{\rm DM}$ to levels well below the observed one, especially for 
$2 \, r_\ell^{-1} \sim n \, R_q^{-1} \, (n \in \mathbb{Z}^+)$, namely regions
of parameter space where the aforementioned $s$-channel processes are 
relatively close to being on resonance. 

This curious effect has profound implications, with multi-TeV DM being
quite in consonance with all bounds (relic density, dedicated direct
and indirect searches as well as the LHC constraints), without any
fine tuning being necessary. In addition, this scenario promises very
interesting signals at colliders (both in the forthcoming runs of the
LHC as well as in future high-energy $e^+e^-$ colliders). It is, thus,
worthwhile, to consider such scenarios very seriously and subject them to 
rigorous tests, both in the context of colliders as well as low-energy 
constraints, {especially those pertaining to flavour. We plan to examine 
real issues in future publication.}

\appendix
\section{Appendix : One loop contributions to gauge boson self energies}
In calculating the loop-contributions of the KK-particles to
$\mathcal{S,T}$, we consider only the leading terms and neglect
effects due to the deformations of the wavefunctions.
It is useful, in this context to define the function
\beq
f(m_n,m_1,m_2) = \log \frac{m_n^2-x(1-x)p^2+(1-x)m_1^2+x\,m_2^2}{\mu^2}
\eeq
where the dependence of $f$ on the momentum $p$ and the renormalization 
scale $\mu$ has been suppressed. The dependence of the measurables 
$\mathcal{S,T}$ on $\mu$ actually vanishes, as it should. 

We, now, list the various loop-contributions to the self-energies in 
question. We use dimensional regularization, working in $4 - \epsilon$ 
dimensions and define
\beq
{\cal A} \equiv \frac{2}{\epsilon} \, - \gamma_E + \log (4\pi)
\eeq
where $\gamma_E$ is the Euler-Mascheroni constant.

\noindent 
\underline{Contribution of KK quarks}

Numerically, only the contribution due to the top-partners are of any
importance and these are given by
\beq
\barr{rcl}
\Pi^{Q}_{\gamma\gamma}(p^2) &=&\dis 
\frac{\alpha}{4\pi}\frac{-8}{3}\int^1_0 dx\,\left[\frac{5}{3}
{\cal A}- 2x(1-x)p^2 \, 
     \left(4 f(m_n,m_t,m_t)+ f(m_n,0,0)\right)\right]
\\[2ex]
\Pi^{Q}_{ZZ}(p^2)&=&\dis 
\frac{\alpha}{4\pi} \, \frac{-3 + 8 s_w^2-\frac{32}{3}s_w^4}{s_w^2c_w^2} \, 
   (2 \, p^2) \, \int^1_0 dx\, x(1-x) \, \left({\cal A}- f(m_n,m_t,m_t)\right)
\\[1ex]
&+&\dis \frac{\alpha}{4\pi}\, \frac{3}{s_w^2c_w^2} \,
\int^1_0 dx\,\left({\cal A}- f(m_n,m_t,m_t)\right)m_t^2 
\\[1ex]
&+&\dis 
\frac{\alpha}{4\pi}\, \frac{-3 + 4 s_w^2-\frac{8}{3}s_w^4}{s_w^2c_w^2} \, 
   (2 \, p^2) \, \int^1_0 dx\,x(1-x) \, \left({\cal A}- f(m_n,0,0)\right)
\\[2ex]
\Pi^{Q}_{Z\gamma}(p^2)&=&\dis 
\frac{\alpha}{4\pi} \, \frac{-4+\frac{32}{3}s_w^2}{s_wc_w} \, 
  (2 \, p^2) \, \int^1_0 dx\, x(1-x) \, \left({\cal A}- f(m_n,m_t,m_t)\right)
\\[1ex]
&+&\dis \frac{\alpha}{4\pi} \, \frac{-2+\frac{8}{3}s_w^2}{s_wc_w} \,
    (2 \, p^2) \, \int^1_0 dx\,x(1-x) \, \left({\cal A}- f(m_n,0,0)\right) 
\\[2ex]
\Pi^{Q}_{WW}(p^2)&=& \dis \frac{\alpha}{4\pi} \, \frac{-6}{s_w^2}\,
     \int^1_0 dx\,\left({\cal A}
- f(m_n,m_b,m_t)\right)\left(2x(1-x)p^2-x\,m_t^2-(1-x)\,m_b^2\right)\\
\earr
\eeq

\noindent
\underline{Contribution of KK Gauge bosons and Ghosts}

We sum here the contributions of the gauge bosons and ghosts of a 
given KK order. Further, we neglect the small splitting between the various 
fields at a given level emanating from electroweak symmetry breaking. 
Remembering that the KK-idex, $n$ now has two components, 
we have
\beq
\barr{rcl}
\Pi^{GB}_{\gamma\gamma}(p^2)&=&\dis \frac{\alpha}{4\pi}\dis 
\int^1_0 dx\,
\left({\cal A}- f(m_n,m_W,m_W)\right) \, \left((-12x^2+20x-3)p^2+2m_W^2\right) 
\\[2ex]
\Pi^{GB}_{ZZ}(p^2)&=&\dis \frac{\alpha}{4\pi} \, \frac{c_w^2}{s_w^2}
   \, \int^1_0 dx\, \left({\cal A}- f(m_n,m_W,m_W)\right) \, 
     \left((-12x^2+20x-3)p^2+2m_W^2\right) 
\\[2ex]
\Pi^{GB}_{\gamma Z}(p^2)&=&\dis \frac{\alpha}{4\pi} \, \frac{c_w}{s_w} \,
   \int^1_0 dx\,  \left({\cal A}- f(m_n,m_W,m_W)\right) \, 
       \left((-12x^2+20x-3)p^2+2m_W^2\right)
\\[2ex]
\Pi^{GB}_{WW}(p^2)&=&\dis \frac{\alpha}{4\pi} \, \frac{c_w^2}{s_w^2} \,
\int^1_0 dx\, \left({\cal A}- f(m_n,m_W,m_Z)\right) \, 
\\
&&  \dis    \hspace*{8em}  \left[ (-12x^2+12x+1)p^2+2(2-3x)m_Z^2 + 2(3x-1)m_W^2\right]
\\[1ex]
&+& \dis \frac{\alpha}{4\pi}\int^1_0 dx\,\left({\cal A}- f(m_n,m_W,0)\right)\,
\left[(-12x^2+12x+1) p^2 + 2(3x-1)m_W^2\right]
\earr
\eeq

\noindent \underline{Contribution of KK Higgs bosons}

For the case of Higgs bosons defined on the 4-branes at $x_4 = 0, \pi$, 
we have 
\beq
\barr{rcl}
\Pi^{H}_{\gamma\gamma}(p^2)&=& \dis 
\frac{\alpha}{4\pi}\, \int^1_0 dx\, \left({\cal A}- f(m_n,m_W,m_W)\right) \, 
\left((-4x^2+6x-2)p^2-2m_W^2\right) 
\\[2ex]
\Pi^{H}_{\gamma Z}(p^2)&=&\dis \frac{\alpha}{4\pi} \, \frac{s_w}{c_w}\, \int^1_0 dx\,
\left({\cal A}- f(m_n,m_W,m_W)\right) \, \left((2x^2-3x+1)p^2+2m_W^2\right) 
\\[2ex]
&+& \dis \frac{\alpha}{4\pi} \, \frac{c_w}{s_w}\, \int^1_0 dx\,
        \left({\cal A}- f(m_n,m_W,m_W)\right) \, \left((-2x^2+3x-1)p^2\right) 
\\[2ex]
\Pi^{H}_{Z Z}(p^2)&=&\dis \frac{\alpha}{4\pi} \, \frac{1}{4\,c_w^2\,s_w^2}\, \int^1_0 dx\,
\left({\cal A}- f(m_n,m_Z,m_H)\right) 
\\
& & \dis \hspace*{8em} \left[ (-4x^2+4x-1)p^2 + (1-2x)m_H^2+(2x-5)m_Z^2\right]
\\[2ex]
& +& \dis \frac{\alpha}{4\pi} \, \frac{c_{2w}^2}{2\,s_w\,c_w}\, p^2 \, \int^1_0 dx\,
\left({\cal A}- f(m_n,m_W,m_W)\right)  \, (-2x^2+3x-1)
\\[1ex]
&-& \dis \frac{\alpha}{2\pi}s_w^2\, m_Z^2  \, \int^1_0 dx\,
\left({\cal A}- f(m_n,m_W,m_W)\right)
\\[2ex]
\Pi^{H}_{WW}(p^2)&=&\dis
\frac{\alpha}{4\pi} \, \frac{1}{4s_w^2}\, \int^1_0 dx\,
\left({\cal A}- f(m_n,m_W,m_Z)\right)
\\
& & \dis \hspace*{8em} \left[ ((4x-4x^2-1)p^2+(1-2x)m_Z^2 + (2x-1)m_W^2\right]
\\[1ex]
& -& \dis \frac{\alpha\,s_w^2}{4\pi}\, \int^1_0 dx\,
\left({\cal A}- f(m_n,m_W,m_Z)\right) \, m_Z^2
\\[1ex]
&-&\frac{\alpha}{4\pi}\, \int^1_0 dx\,\left({\cal A}-
 f(m_n,m_W,m_H)\right) 
\\
& & \dis \hspace*{8em} \left[ (4x-4x^2-1)p^2+(2x-5)m_W^2+ (1-2x)m_H^2\right]
\earr
\eeq

\section*{Acknowledgement}
The authors would like to thank Alexander Pukhov for discussions in regards to micrOMEGAs.
 MTA acknowledges the support from the SERB National Postdoctoral
fellowship [PDF/2017/001350].  DS would like to thank UGC-CSIR, India
for financial assistance.

\bibliographystyle{JHEP}
\bibliography{reference}
\end{document}